# Estimation of genomic characteristics by analyzing k-mer frequency in *de novo* genome projects

**Binghang Liu[1,2]*, Yujian Shi[1]*, Jianying Yuan[1]*, Xuesong Hu[1,3], Hao Zhang[1], Nan Li[1], Zhenyu Li[1], Yanxiang Chen[1], Desheng Mu[1], Wei Fan[1,3]***

[1]BGI-Shenzhen, Shenzhen, 518083, China.

[2]HKU -BGI Bioinformatics Algorithms and Core Technology Research Laboratory, HongKong

[3]Biodynamic Optical Imaging Center, Peking University, Beijing 100871, China.

*Correspondence should be addressed to:

Binghang Liu: binghang.liu@qq.com

Yujian Shi: shiyujian@genomics.cn

Jianying Yuan: yuanjianying@genomics.cn

Wei Fan: fanweiagis@126.com

## Abstract

**Background**: With the fast development of next generation sequencing technologies, increasing numbers of genomes are being de novo sequenced and assembled. However, most are in fragmental and incomplete draft status, and thus it is often difficult to know the accurate genome size and repeat content. Furthermore, many genomes are highly repetitive or heterozygous, posing problems to current assemblers utilizing short reads. Therefore, it is necessary to develop efficient assembly-independent methods for accurate estimation of these genomic characteristics.
**Results**: Here we present a framework for modeling the distribution of k-mer frequency from sequencing data and estimating the genomic characteristics such as genome size, repeat structure and heterozygous rate. By introducing novel techniques of k-mer individuals, float precision estimation, and proper treatment of sequencing error and coverage bias, the estimation accuracy of our method is significantly improved over existing methods. We also studied how the various genomic and sequencing characteristics affect the estimation accuracy using simulated sequencing data, and discussed the limitations on applying our method to real sequencing data.
**Conclusion**: Based on this research, we show that the k-mer frequency analysis can be used as a general and assembly-independent method for estimating genomic characteristics, which can improve our understanding of a species genome, help design the sequencing strategy of genome projects, and guide the development of assembly algorithms. The programs developed in this research are written using C/C++, and freely accessible at Github URL (https://github.com/fanagislab/GCE) or BGI ftp ( ftp://ftp.genomics.org.cn/pub/gce).

## Background

In recent years, more and more large genomes have been assembled by Whole-Genome-Shotgun (WGS) short reads generated from next generation sequencing (NGS) technologies [1], including the panda [2], potato[3], and many others. Many genomes are highly repetitive [4] or polyploid [5], and some species have highly heterozygous diploid genomes [6]. These genomic characteristics will increase the difficulty of the assembly processes, resulting in incomplete and fragmental assembled sequences [7, 8], making it impossible to infer the accurate genome size and repeat content only based on these assembled sequences. Another issue is that when genomes have an extremely-high repeat content or heterozygous rate or are polyploid, the assembled sequences using current available algorithms and short reads may become too fragmental and even unusable, and so it is very important to get an accurate estimation of genomic characteristics before deciding whether or not to start large-scale genome sequencing project.

Several experimental technologies have been developed to explore these genomic characteristics. Feulgen densitometry and flow cytometry are the two most popular techniques used to estimate the genome size, which is presented as the C-value [9]. DNA reassociation kinetics, also known as $C_0t$ analysis, is usually used to measure and classify the repetitive DNA sequences in a genome [10]. Some previous studies have performed estimation of heterozygosity using molecular markers [11] or DNA microarrays [12], however, the performance of these techniques is often poor.

In *de novo* genome projects, analyzing the k-mer frequency, which is independent of genome assembly, is widely used as an alternative way to estimate the genome size [2, 3, 13, 14]. However, most projects have adopted a very rough estimation method (denoted below as “integer precision estimation”), and their estimations are often not so accurate. The group of Michael S. Waterman was the first to perform systematic study on the estimation of the genome size and repeat structure using k-mer frequency, and in 2003 they published a basic estimation method using the Mixed-Poisson Model and EM algorithm [15]. Built on theoretically perfect data, their method has many limitations, and the authors did not provide a usable tool for real application. In 2009, Shan & Zheng extended the Michael S. Waterman's method, and published a more generally applicable software for genome size prediction (GSP) using the Bayesian estimation (BE) and EM iteration [16]. GSP was initially developed for perfect data, and then adopts a simple approach to tolerate some sequencing errors. Due to the quite complex characteristics of real sequencing data and the immature status of the two methods, neither of them has been widely adopted by real applications till now. Besides from them, we can't find other more formal publications on this issue. In fact, it is still very difficult for non-bioinformaticians to estimate the genomic characteristics properly.

In this article, we developed a more sophisticated method to satisfy the demands of real applications. We improved the estimation accuracy over existing methods by introducing the k-mer individuals and float precision estimation technique, and our method fully considers sequencing characteristics such as error and coverage bias and has integrated a module for processing them in order to get the best estimation of genomic characteristics. Futhermore, we extend the previous application of estimating genome size and repeat structure to heterozygous rate estimation. We also studied how the various genomic and sequencing characteristics affect the estimation accuracy with simulated data, and demonstrated the application potential and limitations of our model using real sequencing data from several finished genome projects. We hope this work can help more genome projects to estimate the genomic characteristics more accurately, and thus assist the understanding of their genome biology. We also suggest that future genome projects, for which these genomic characteristics are not clear and may potentially pose serious problems in assembly, to initially perform some small-scale sequencing (5~25X) and estimate the genomic characteristics by k-mer frequency, which can then help determine the best large-scale sequencing strategies and most suitable assembly algorithms.

# Methods

## Counting k-mer frequency

The counting of k-mer frequency in the sequencing data can be carried out using many of the currently available tools, such as Meryl [17], Tallymer [18], and jellyfish [19], here we use our in-house software Kmerfreq. Note that before counting, k-mer size ($K$) needs to be determined. $K$ should be kept small to prevent the overuse of computer memory, while still large enough so that most k-mers are unique in the genome. Once $K$ is determined, the maximum number of k-mers is fixed as $4^K$. More detailed methods are shown in Supplemental materials.

## Estimating sequencing depth and genome size

As an easy-to-understand illustrative example, we will first discuss the simplest k-mer frequency (depth) model using hypothetical "ideal" reference genome and sequencing data. To start with, we have to introduce two concepts: genomic frequency and coverage depth, which are used to refer to the k-mer frequency counted from reference genome and sequencing data respectively. The "ideal" reference genome here is assumed to be a random sequence, with no heterozygosity and no repeats for a certain k-mer size, meaning the genomic frequency for all of these k-mers is 1. The "ideal" sequencing data here is assumed to be produced from randomly single-ended and equal-length whole genome shotgun process [20] without any sequencing errors or coverage bias, such data meaning that the distribution for the start positions of reads follows a Poisson distribution. When the read length ($L$) is far shorter than the genome size ($L<<G$), the bases and k-mers can be also thought to be generated by random processes and their coverage depth will also follow Poisson distributions [15, 21] (Figure 1a). Based on Poisson theory as well as the relationship between base number and k-mer number, the sequencing depth (expected base coverage depth) and genome size can be calculated by formulas (1) and (2) shown below[15], both of which are important parameters for *de novo* projects.

Let $n_{base}$, $n_{k\text{-}mer}$ be the total number of bases and k-mers from sequencing data, and $c_{base}$, $c_{k\text{-}mer}$ be the expected coverage depth for bases and k-mers, then we can get $c_{base} = n_{base} / G$ and $c_{k\text{-}mer} = n_{k\text{-}mer} / G$. As one read with length $L$ generates $L-K+1$ k-mers, $n_{k\text{-}mer} / n_{base} = (L-K+1) / L$. Thus:

$$c_{base} = c_{k\text{-}mer} \times L / (L - K + 1) \quad (1)$$

$$G = n_{k-mer} / c_{k-mer} = n_{base} / c_{base} \quad (2)$$

Note: Formula (1) and (2) were firstly deduced by Michael S. Waterman's group [15], and we list them here to help understanding of the bellowing formulas. For simplification, in the following parts of this paper, we will exclusively use $c$ to represent $c_{k\text{-}mer}$, and $n$ to represent $n_{k\text{-}mer}$.

Given that the read length ($L$) and k-mer size ($K$) are fixed values, to accurately estimate the sequencing depth and genome size using the above formulas (1) and (2), we must firstly determine two parameters: the total number of k-mers ($n$) and the expected k-mer coverage depth ($c$). The $n$ parameter can be directly obtained from the k-mer counting results. The parameter $c$ is the key parameter to be estimated in this part [15, 16], and it can be inferred from the widely adopted Poisson distribution of k-mer frequency, denoted here as the k-mer species curve (Figure 1a), represented in formula (3), which fits well-known Waterman's estimation method [15]. Here we also introduce another equivalent curve, the k-mer individuals curve (Figure 1a), represented in formula (4). The number of k-mer individuals is the product of k-mer species number and corresponding depth value. The points on these two types of k-mer coverage depth curves indicate the ratio of k-mer species or individuals classified by each k-mer depth value respectively.From observation on Figure 1a and deduction in formula (4), we found that the k-mer individuals curve is a variation of Poisson distribution (denoted as varied-Poisson), which has the same figure shape but moves rightwards wholly by one unit. In practice, we can use either of these two curves to estimate the $c$ value, or combine the results from each curve to make more solid estimation. The two basic models shown in formulas (3) and (4) were denoted as the "basic model" in this paper.

Probability density function of k-mer species curve for the "ideal" genome:

$$P_{Kspecies}(x) = \frac{c^x}{x!} e^{-c} \tag{3}$$

Probability density function of k-mer individuals curve for the "ideal" genome:

$$P_{Kindividuals}(x) = xP_{Kspecies}(x) / c = P_{Kspecies}(x-1) \tag{4}$$

Note: Formula (3) was first introduced by Michael S. Waterman's group [15], while formula (4) was introduced by us for the first time. $x$ in the two probability density functions, refer to the k-mer coverage depth.

After we obtained one of the two k-mer coverage depth curves, a quick and rough way is to use the observed peak depth value as the estimated $c$, which is an integer value. This method is denoted as integer precision estimation here, which has been widely adopted by previous and current genome projects. However, in most cases, the real $c$ value is not an integer (Figure 1b). To estimate c more accurately with float precision, we developed a new algorithm by using the relationships between the neighboring points in the k-mer coverage depth distribution curve, denoted as float precision estimation here, shown as the formulas (5) and (6). The detailed deductions are shown in Supplemental materials.

Formulas to calculate $c$ with float precision:

$$c = \frac{P_{Kspecies}(x+1)}{P_{Kspecies}(x)}(x+1) \tag{5}$$

$$c = \frac{P_{Kindividuals}(x+1)}{P_{Kindividuals}(x)} x \tag{6}$$

Note: Theoretically, x is arbitrary, and a pair of neighboring points is enough for estimation of c. However, to estimate c more accurately, it is better to calculate c independently using 5 to 10 pairs of points adjacent to the depth peak and adopt the average, to our experience.

## Exploring repetitive genomes

We used the "ideal" genome to introduce the basic model above, but real genomes often contain differing amounts of repeat sequences [22], which bring great challenges to the assembly processes [23]. In this section, we will use the human reference genome and the "ideal" sequencing data simulated from it as an example to explore the relationship between k-mers and repeats.

The k-mers localized in repeat regions will not appear uniquely in the genome, and their genomic frequencies are in line with the copy number of the repeats. We firstly counted the genomic k-mer ($K$=17) frequency in the human reference genome. Then k-mers were classified by each genomic frequency ($i$), and the ratio of k-mer species ($a_i$) and individuals ($b_i$) for each class were calculated (Figure 2a). We also plotted the k-mer species and individuals curves from the simulated sequencing data shown in Figure 2b. Repeats in the genome will cause several peaks on both the k-mer species and individuals curves, the heights of which are closely related to the $a_i$ and $b_i$ values respectively in Figure 2a. The k-mer species curve for repetitive genome had previously been modeled as a compound of discrete Poisson distributions [15], shown in formula (7), each of which was generated by a class of k-mers with specific genomic frequency ($i$) and expected coverage depth ($c_i$) [15]. We showed that the k-mer individuals curve in Figure 2b can also be modeled as a compound of discrete varied-Poisson distributions in similar way, shown in formula (8). Note that these two models are denoted as the "standard model" in this paper.

Assuming the range of genomic k-mer frequency is [1,m], $a_i=n_{i,genomic,Kspecies}/n_{genomic,Kspecies}$ and $b_i=n_{i,genomic,Kindividuals}/n_{genomic,Kindividuals}$ are the ratios of k-mer species and individuals with genomic frequency $i$, and $c_i=i\times c$ is the expected coverage depth of k-mers with genomic frequency $i$ ,where $c$ is the expected coverage depth of unique k-mers (genomic frequency $i$=1). Then we can get:

Probability density function of k-mer species curve for repetitive genome:

$$P_{Kspecies}(x)=\sum_{i=1}^{m} a_i \times P_{Kspecies,i}(x) \qquad (7)$$

Probability density function of k-mer individuals curve for repetitive genome:

$$P_{Kindividuals}(x)=\sum_{i=1}^{m} b_i \times P_{Kindividuals,i}(x) \qquad (8)$$

Note: $P_{Kspecies,i}(x)$ means Poisson distribution with expected coverage depth $c_i$, $P_{Kindividuals,i}(x)$ means varied-Poisson distribution with expected coverage depth $c_i$.

Formula (7) was first introduced by Michael S. Waterman's group [15], we listed it here for comparison with formula (8).

Both of the $a_i$ and $b_i$ values can reflect the repeat structure in genome, though they have slightly different meaning. For simplification, we only use $a_i$ to describe repeats, because $a_i$ and $b_i$ values can be converted to each other. To estimate the $a_i$ values for *de novo* projects where a reference genome does not exist, the Waterman [15] and Shan [16] groups have provided alternative EM approaches based on Bayes models, in which each k-mer species has two attributes: the genomic frequency and the coverage depth. Firstly, the experience-based or equal prior probability values are assigned to all the genomic k-mer frequencies, and then the posterior probability values for these genomic k-mer frequencies are calculated using formula (9) shown below. Next, the resulting posterior probability values are used as the input prior probability and this process is iterated until the input prior and resulting posterior probabilities are merged. During these iteration cycles, the $c_i$ values are also adjusted to be more accurate.

Here we used a similar method to Waterman [15] and Shan [16]groups for $a_i$ estimation, but a different method for c estimation, shown in formula (10) . Essentially, this method is equivalent to applying our float precision estimation method shown in formula (5) on the part of genomic unique k-mers (genomic frequency == one). These genomic unique k-mers are obtained by removing the contributions of repeat sequences (genomic frequency >= 2) on the major peak region of the k-mer species curve, with the help of estimated $a_i$ values whose accuracy are improved along with the Bayes iteration cycles. Here the major peak should be formed by the genomic unique k-mers. If the difference between the major and minor peaks is not significant, larger k-mer size should be used, shown in Figure 2cd. Based on the compound Poisson model, we found that the genome size can be calculated by the total k-mer number ($n_{k\text{-}mer}$) and the expected coverage depth of unique k-mers ($c$): $G = n_{k\text{-}mer} / c$. Using the iterated re-calculated more accurate $c$, we can estimate genome size with higher accuracy. More detailed deductions are shown in Supplemental materials

In the Bayes model, we define:

$a_i$: the ratio (probability) of k-mer species with genomic frequency $i$, $i$=1,2,….,m.

$v_j$: the ratio (probability) of k-mer species with coverage depth $j$, $j$=0, 1, 2, …,w.

$c$: the expected coverage depth for unique k-mer class with genomic frequency of one.

The prior probabilities:

$$P(i) = a_i, P(j \mid i) = \frac{(ic)^j e^{-ic}}{j!}$$

$$P(i \mid j) = \frac{P(i)P(j \mid i)}{\sum_{i=1}^{m} P(i)P(j \mid i)} = \frac{a_i P(j \mid i)}{\sum_{i=1}^{m} a_i P(j \mid i)}$$

The posterior probabilities: $a_i = \sum_{j=0}^{w} P(i \mid j) \times v_j$

Therefore giving the iteration formula:

$$a_{i,t+1} = \sum_{j=0}^{w} \frac{a_{i,t} P(j \mid i)}{\sum_{i=1}^{m} a_{i,t} P(j \mid i)} \times v_j \qquad (9)$$

$$c_{t+1} = \frac{P_{Kspecies,1,t+1}(x+1)}{P_{Kspecies,1,t+1}(x)} \times (x+1) \qquad (10)$$

Note: $P(i)$ is the probability of one k-mer species with genomic frequency $i$. $P(j/i)$ is the probability of one k-mer species with genomic frequency $i$ to be covered $j$ times in sequencing data. Although it is not possible to count $v_0$, which would be the ratio of no covered k-mer species, we can calculate it by $v_0 = \sum a_i \times P_{Kspecies,i}(0)$, where $t$ refer to the iteration cycles. The $c$ value is re-calculated by the float precision estimation method using the major peak formed by unique k-mer class with a genomic frequency of one. As the estimation of $a_i$ gets more accurate, the $c$ value will

also become more accurate.

## Exploring the heterozygous genomes

Besides repeats, most diploid genomes are heterozygous to a different degree, and most current de bruijn graph (DBG) based assemblers cannot easily handle NGS short reads derived from highly heterozygous genomes. It is very useful to estimate the heterozygous rate in the early stage of a genome project. The Waterman and Shan groups have not addressed the issue of heterozygous rate estimation.

To simplify things, the heterozygosity discussed in this paper is restricted to SNPs. We firstly simulated SNP sites randomly in the "ideal" reference genome (haploid) to create "ideal" heterozygous genome (diploid) and simulated "ideal" sequencing data from it. Then the k-mers can be divided into 2 classes: the heterozygous k-mers and homozygous k-mers, which are generated by the heterozygous and homozygous genomic regions respectively. It is supposed that when the heterozygous rate is relatively low, the SNP sites will be distributed sparsely in the whole genome, so there will be ideally 2 $\times$K heterozygous k-mers around each SNP site. These heterozygous k-mers have half of the expected coverage depth and cause a new peak at 1/2 expected coverage depth in both the k-mer species and k-mer individuals curves shown in Figure 3ab. In order to be consistent, the k-mer genomic frequency for heterozygous genome here is counted using the two haploid genomes and then dividing them by 2, then for the "ideal" heterozygous genome here, there are only two genomic frequency values 1/2 and 1, and the heterozygous rate can be roughly estimated with $a_{1/2}$ using formula(11).

For heterozygous repetitive genome, the number of k-mer genomic frequency values will be doubled compared to non-heterozygous repetitive genome, so there will also be 1.5, 2, 2.5, 3, etc. The k-mer species and individuals curves for repetitive genomes (human) with various heterozygous rate are shown in Figure 3cd. Theoretically, the number of peaks will also be doubled compared to non-heterozygous repetitive genome. We extend the standard model mentioned above to model heterozygosity, shown in formulas (12) and (13), and denoted as the "heterozygous model". In this model, there is no change in the formulas to estimate $c$ and $G$, as well as the Bayes model to estimate the $a_i$ values except for the step of $i$ ,which changes from integers to half of the integers for diploid genomes.

Because $c$ should be estimated from the major peak generated by genomic homozygous unique k-mers, the key problem for genome size estimation of heterozygous genome is to distinguish the peaks on k-mer species or individuals curve. In practice, we can determine the homozygous unique peak with the aids of some biological information or the rough assembly result. Though the heterozygous rate for non-repetitive genome can be estimated by formula (11), it cannot be extended to repetitive genome, because that there is no clear relationship between the $a_{1/2}$ and heterozygous rate here. More detailed methods are shown in Supplemental materials.

Formulas for heterozygous genomes:

$$\eta = \frac{a_{1/2} n_{Kspecies} / (2K)}{n_{Kspecies} - a_{1/2} n_{Kspecies} / 2} = \frac{a_{1/2}}{K(2 - a_{1/2})} \quad (11)$$

$$P_{Kspecies}(x) = \sum_{i=1/2}^{m(step=1/2)} a_i \times P_{Kspecies,i}(x) \quad (12)$$

$$P_{Kindividuals}(x) = \sum_{i=1/2}^{m(step=1/2)} b_i \times P_{Kindividuals,i}(x) \quad (13)$$

### Dealing with sequencing error and coverage bias

The current real sequencing data usually have sequencing error and coverage bias problems, and so the k-mer species or individuals curve reflects both the genomic and sequencing characteristics, the mixture of which makes it difficult to estimate each type of characteristics accurately. Shan's group [16] has adopted a simple method to process sequencing errors, but neither Waterman's nor Shan's groups has addressed the coverage bias problems. In the following, we will show how sequencing error and coverage bias can affect the k-mer species and individuals curves, and point out methods to deal with each of them in order to improve the estimation of genomic characteristics.

Previous studies have demonstrated the k-mer distribution characteristics caused by sequencing errors [16, 19]. To be more clear, here we illustrated this problem on Figure 4abc. Sequencing error will generate a sharp left-side peak on the k-mer species curve (Figure 4a). As the rate of sequencing errors increases, the left-side peak becomes larger, meanwhile, other normal peaks in the curve would become smaller and also moves leftwards (Figure 4bc). Most of erroneous k-mers caused by sequencing error are non-genomic k-mers, which appear in very low frequency and form the left-side sharp peak. The frequency of the left erroneous k-mers, which have the same sequence with the genomic k-mers, will be merged with the correct k-mers. Then it is impossible to distinguish them just by frequency (Figure 4a).

As the distribution of erroneous k-mer is so complex, in practice, we use a simple method to deal with sequencing error adapted from that of Shan's group [16]. They just exclude the low depth k-mers in the left-side sharp peak with a threshold at the lowest turning point (Figure 4a), and take the left high depth k-mers as correct k-mers, which are used for estimating the $c$ and $a_i$ values. Using this method, most of the erroneous k-mers have been removed, but certain number of correct k-mers will also be removed (Figure 4a), which will reduce the estimation accuracy. In our method, we additionally use the estimated $c$ and $a_i$ values to rebuild the theoretic distribution curve, and compare it with the real distribution curve to adjust the number of correct k-mers.

Coverage bias [24-26] is a more complex issue than sequencing error, and there is few publications discussed its affection on k-mer distribution characteristics. By analyzing genomic data from the deadly 2011 German *E. coli* 0104:H4 outbreak [27], we found that there may be many other factors besides GC content that can contribute to the coverage bias (Figure 4d). Coverage bias will flatten the k-mer species or individuals curve, making it more difficult to observe the peak depth clearly and estimate $c$ accurately. Because of coverage bias, the probabilities of sampling k-mers in the same genomic frequency class are not equal but appear in a continuous spectrum, so the standard and heterozygous models with discrete Poisson distributions need further development. This problem has not been previously considered in the published models. Here we use an extended form with continuous Poisson or varied-Poisson distributions to model the k-mer distribution with coverage bias, as shown in formulas (14) and (15), and denoted as the "continuous model". In theory, $\tilde{i}$ does not mean genomic frequency any more, and so $\tilde{a}_{\tilde{i}}$ does not reflect ratio of genomic frequency either, but it is related to the sampling probabilities of k-mers in the sequencing data.

As a practical step, we used dense ranks of $\tilde{i}$ with either a fixed equal step or dynamic non-equal step to simulate the continuous model. If we set a fixed equal step of $\tilde{i}$, such as 1/4, 1/8, 1/16, the continuous model is just a further extension of the heterozygous model in which the step of $i$ is 1/2. Here we suggest using the dynamic non-equal step model, which has advantage in both accuracy and convenience, as shown in formulas (16) and (17). In this model, we use character "*k*" to replace "$\tilde{i}$" because of the dynamic non-equal step. The estimation of $\tilde{a}_k$ in the Bayes iteration is similar with that of fixed equal step models such as the standard and heterozygous models, but the estimation of $\tilde{c}_k$ is different, as shown in formulas (18) and (19) respectively. Based on our experiences demonstrated in Supplemental materials, we take the $\tilde{c}_k$ with the highest $\tilde{a}_k$ value as the $c$ value, which was used to estimate the genome size. Although it is difficult to infer genomic $a_i$ values from the estimated $\tilde{a}_k$ values, we can sum up the multiple $\tilde{a}_k$ values around each peak to roughly reflect each genomic $a_i$ values, especially for $a_1$.

Probability function for k-mer species and individuals curves with coverage bias:

$$P_{Kspecies}(x) = \int \tilde{a}_{\tilde{i}} \times P_{Kspecies,\tilde{i}}(x) d\tilde{i} \quad (14)$$

$$P_{Kindividuals}(x) = \int \tilde{b}_{\tilde{i}} \times P_{Kindividuals,\tilde{i}}(x) d\tilde{i} \quad (15)$$

Implementing this, we used dense ranks with dynamic non-equal steps to simulate the continuous model:

$$P_{Kspecies}(x) = \sum_{k=1}^{m'} \tilde{a}_k \times P_{Kspecies,k}(x) \quad (16)$$

$$P_{Kindividuals}(x) = \sum_{k=1}^{m'} \tilde{b}_k \times P_{Kindividuals,k}(x) \quad (17)$$

Note: $m'$ is the total number of Poisson distributions considered. The symbol $k$ here stands for the order, and there is no relation between $k$ and $\tilde{c}_k$.

The formulas in Bayes iterations for this dense discrete model:

$$P(j \mid k) = \frac{\tilde{c}_k^{\,j} e^{-\tilde{c}_k}}{j!}$$

$$\tilde{a}_{k,t+1} = \sum_{j=0}^{w} \left[ \frac{\tilde{a}_{k,t} P(j \mid k)}{\sum_{k=1}^{m'} \tilde{a}_{k,t} P(j \mid k)} \times v_j \right] \quad (18)$$

$$\tilde{c}_{k,t+1} = \frac{1}{\tilde{a}_{k,t+1}} \sum_{j=0}^{w} \left[ \frac{\tilde{a}_{k,t+1} P(j \mid k)}{\sum_{k=1}^{m'} \tilde{a}_{k,t+1} P(j \mid k)} \times v_j \right] \quad (19)$$

Note: the formula to calculate $\tilde{c}_k$ here is similar to that of the Waterman group.

# Results and discussion

## Relationship between data coverage and estimation accuracy

In this section, we investigate how data coverage affects the accuracy of genome size estimation. To simplify matters, we will still use the "ideal" reference genome and simulated sequencing data as an example. We evaluated the accuracy of estimation using the ratio between deviation size and real genome size ($\Delta G/G$). The $\Delta G/G$ values calculated by the 4 different methods (k-mer species and individuals, together with integer precision estimation and float precision estimation) with various expected k-mer coverage depth were plotted in Figure 5, and detailed methods to calculate $\Delta G/G$ values are shown in Supplemental materials. The two integer precision ways give similar results, ranging from 0.1%-10%; whist the two float precision ways also show similar results, ranging from 0.01%-0.1%. When the expected k-mer coverage depth ($c_{k\text{-}mer}$) is lower than 20, there is a roughly two orders of magnitude difference between the integer precision and float precision methods; when $c_{k\text{-}mer}$ is higher than 20 but lower than 85, there is still roughly a one order of magnitude difference, which illustrate the significant advantage for the float precision estimations over the integer precision estimations, especially when the $c_{k\text{-}mer}$ is relatively low. It should be noted that the $\Delta G/G$ values from integer precision methods appear in periodic character, and in particular the estimated genome size is almost always larger than the real value. In contrast, the $\Delta G/G$ values from float precision ways are randomly distributed, i.e. no systematic bias, which indicates the potential of getting the accurate estimation by averaging the results from multiple experiments.

## Estimating with various types of simulated data

In order to evaluate the performance of our models and analyze the influence of repeat content,

heterozygous rate, sequencing error and k-mer size on the accuracy of genomic characteristics estimation, we simulated 48 sets of 25X coverage data from 4 species (*E. coli*, *Arabidopsis*, Human, and Maize, the reference genome versions are shown in Table 1), with 4 ranks of heterozygous rates (0%, 0.01%, 0.1%, 1%) and in combination with 3 ranks of sequencing error rate (0%, 0.05%, 1%). As the sequencing coverage bias is too complex to model and simulate, it was not considered in this section. For each data set, we calculated the 17-mer and 25-mer $a_i$ values from the reference genomes. The reference genome sizes and some major $a_i$ values are shown in Table 1, and the complete information of $a_i$ can be found in Table S1. The k-mer species and individuals curves were plotted for each data set, and are shown in Figure S1.

To estimate the genome size and $a_i$ values from the simulated sequencing data, we applied the heterozygous model to data sets with a heterozygous rate equal or larger than 0.1% and the standard model to the other data sets, due to the fact that the heterozygous model is not particularly sensitive with extreme-low heterozygous rate. We used $\Delta G/G$ to evaluate the accuracy of genome size estimation. The distribution of $\Delta G/G$ values for maize is shown in Figure 6ab, and those for other species can be found in Figure S2. Within these simulated data sets, about 15%, 73% and 95% of the $\Delta G/G$ values are smaller than 0.1%, 1%, and 5% respectively. For simplicity, we used $\Delta a_1/a_1$ to evaluate the accuracy of $a_i$ estimation, and the distributions of $\Delta a_1/a_1$ values for all species are shown in Figure S3. Similarly to the accuracy level of $\Delta G/G$ values, about 13%, 54%, and 92% of the $\Delta a_1/a_1$ values are smaller than 0.1%, 1%, and 5% respectively. With the estimated $c$ and $a_i$ values, we also calculated the estimated k-mer species and individuals curves using the compound Poisson models, as shown in Figure S1, which are well consistent with their theoretical curves, indicating the high accuracy of $c$ and $a_i$ estimation.

For genome size estimation, we found that higher rates of repeats, heterozygosity and sequencing error would result in lower accuracy, especially when the 3 factors occur together (Figure 6ab, Figure S2 and Figure S4). Among these factors, sequencing error is the most difficult factor to deal with, with the repeat and heterozygosity levels enhancing this effect. Besides these issues, the k-mer size also influences the estimation accuracy. By comparing $\Delta G/G$ values from 17-mer and 25-mer (as shown in Figure 6ab, Figure S5), we suggest using smaller k-mer size for data with high heterozygosity and sequencing error rates. The evaluation of $a_i$ estimation and the analysis of influencing factors are more complex than that of genome size, so it is difficult to get a clear understanding of these. Additional results and discussion relating to this are shown in Supplemental materials.

For repeat structure estimation, we additionally analyzed the non-heterozygous data for the four species using the standard model and k-mer sizes ranging from 11 to 31, as the absolute $a_i$ values are closely related with the k-mer sizes. Although in theory all the $a_i$ and $b_i$ values can be calculated, in practice, we are usually only interested in $a_1$ and $b_1$, which related to the ratio of unique k-mer species in the genome and the ratio of the genome covered by unique k-mers respectively. Both the theoretical and estimated $a_1$ ($b_1$) values with various k-mer sizes are shown in Figure 6cd, and are consistent with each other, indicating a very high accuracy of estimation. As most short read assemblers have adopted the De Bruijn Graph algorithm, obtaining the $a_1$ and $b_1$ values with various k-mer sizes from the raw sequencing data, will be quite helpful for deciding the suitable k-mer size in *de novo* assembly [28].

For heterozygosity estimation, we show all the estimated $a_{1/2}$ values using the heterozygous model and the inferred heterozygous rates by formula (11) in Table S2. We found that the order of magnitude of estimated heterozygous rate is in accordance with that of heterozygous rate we simulated, except for extremely low-heterozygous rates (nearly homozygosis). The estimation accuracy is affected by the repeat content and sequencing error rate, especially when the theoretic heterozygous rate is very low.

## Analyzing real sequencing data from *de novo* genome projects

In this section, we evaluated the performance of our models on real sequencing data from 5 finished *de novo* genome projects, including *E. coli*-O104:H4 [27], the Leaf-cutting ant [14], Potato [3], Panda[2], and YH genome (human diploid reference genome from an anonymous Asian individual) [28, 29]. The repeat content of these genomes is different, and they are nearly homozygous except the panda and YH genomes with a roughly 0.1% heterozygous rate. As the sequencing error rate for real data is usually high and difficult to estimate, on top of counting all k-mers from the raw data, we also adopted two alternative ways to pre-process sequencing error. The first is by ignoring the low quality k-mers in the step of counting k-mer frequency, detailed methods of which can be found in Supplemental materials. The other method is running an error correction tool [28, 30] on the raw reads and then counting k-mer frequency on the error corrected sequencing data.

To estimate the genomic characteristics from these sequencing data, all of which have coverage bias problem, we adopted 3 alternative methods: (1) Rough estimation of genome size, by observing the integer $c$ from the major peak depth and excluding k-mers with depth lower than the lowest-point threshold to obtain the correct k-mer number. (2) Application of the standard model, the float-point $c$ and $a_i$ values can be estimated. As the estimated distribution curve differs greatly with that of real data (Figure 7a), we use the same approach in method (1) to estimate correct k-mer number. (3) Application of the continuous model, the float-point $c$ and $a_i$ values can be estimated, and the estimated distribution curve is well consistent with that of real data (Figure 7b), and we used it to adjust the correct k-mer number. The estimated genome size values for each data set with each method are shown in Table 2, and the estimated information relating to the $a_i$ values can be found in Table S3 and Table S4. Moreover, all the real and estimated k-mer species and individuals curves are shown in Figure S6.

Taking the reported genome sizes in the published papers as reference, we found that over 80% of the ΔG/G or $\Delta a_1/a_1$ values from all the used methods are smaller than 5%, indicating the high estimation accuracy and application potential for real sequencing data. The estimated genome sizes from error corrected data tends to be smaller than that of all raw and low-quality filtered data, which can be explained as that the current error correction tool takes some of the low-frequency k-mer species as erroneous k-mers and removes them in the error corrected data. In contrast, there is no obvious difference between the estimated $G$ values from low-quality filtered data and all raw data, indicating that there is no severe systematical bias in the k-mer filtering process. The estimated genome sizes from the rough estimation tend to be larger than that estimated by standard and continuous models, which may be caused by the low-accurate integer $c$ estimation and the erroneous k-mers. In contrast, there is no obvious difference between the estimated $G$ values from the standard and continuous models, indicating that both methods can be used to estimate the genome size with real sequencing data. Although the standard model is not suitable for data with coverage bias, the peak depth of unique k-mers is almost unchanged, and this model can automatically summarize the effects around each genomic $a_i$ and so it generates similar result to the continuous model. However, there is no obvious trend in the influence of $a_1$ estimation accuracy for either the error processing methods or the estimation models. Additional results and discussions relating to this are shown in Supplemental materials. As the real reference $G$ and $a_i$ values are unknown and the results from different methods fluctuate, it is not easy to determine which method is better, and we feel that further investigation is needed to improve the estimation accuracy with real sequencing data.

To have a clear view of the repeat structure in these genomes as well as choosing suitable k-mer size for the De Brujin Graph assembly, we further estimated the $a_1(b_1)$ values with various k-mer sizes by the continuous model, shown in Figure 7cd. The estimated $a_1(b_1)$ values for smaller k-mer sizes are nearly consistent with the reference values calculated from assembled genomes, but those for larger k-mer sizes become lower than reference values, which may be related to the increasing number and non-uniform distribution of erroneous k-mers. Moreover, the published reference genomes may still be lacking some repeat sequences, resulting in higher reference $a_1(b_1)$ values. Although the estimation accuracy is not as high as that of simulated data, the trend is still correct, and the results can be used to guide *de novo* assembly. Conversely, it is almost impossible to estimate the heterozygous rate for these low heterozygous genomes because of the sequencing coverage bias, which hides the effect of heterozygosity, indicating that the real application of heterozygous rate estimation is limited only to highly heterozygous genomes.

### Comparison with previously published tools

Finally, we compared the performance of our program with published related applications. The Waterman program [15] requires the preparation of input data in a specific format, and this program is only written for testing their model and is unable to be used in practical application, so was not included for comparison. The Shan groups program (GSP)[16] has been designed to predict the genome size using k-mer frequency derived from NGS data, which takes similar input as our program (GCE), so it is convenient for comparisons. To compare the estimation accuracy between these two programs, we simulated two sets of sequencing data from the *E. coli*, *Arabidopsis*, Human and Maize reference genomes, one set containing no sequencing error, with the other set with a 1% sequencing error, and both without heterozygosity, because GSP was not designed for heterozygous rate estimation (although it may have the ability to tolerate some low level of heterozygosity).

The accuracy of estimated genome size and $a_1$ using the two programs GCE and GSP are shown in Table 3. Overall, GCE has a significant higher estimation accuracy over GSP, especially on large and repetitive genomes. Considering on the error-free data first (Table 3a), GSP generates similar results with GCE for relatively repeat-less genomes (*E. coli* and *Arabidopsis*), but generates a poorer result for repetitive genomes (Human and Maize) in comparison to GCE. The explanation for this is likely due to the difference of $c_i$ re-calculation method in the iteration cycles between GSP and GCE.. GCE focused on the unique k-mers ($a_1$), minimized the affection of repeat k-mers and improved estimation accuracy even for the repeat abundant genomes.

We then compared the estimation accuracy using data with 1% sequencing error (Table 3b). The estimation accuracy of genome size and $a_1$ in GCE are both a little lower than that on error-free data. In contrast, the genome size estimation in GSP seems a little better than its error-free results, which may be explained by the fact that GSP tends to generate smaller predictions but sequencing error will make these predictions larger. Notably, the $a_1$ in GSP is worse even for relatively repeat-less genomes. To make a more in-depth investigation, we ran GSP and GCE on the data sets with various combinations of sequencing errors, heterozygous rate, and k-mer sizes, and show the results in Table S5.

## Conclusion

In this article, we have introduced a methodological framework for genomic characteristics analysis based on raw sequencing data, which can estimate genome size, repeat structure and heterozygous rate of the sequenced sample. This is likely not limited solely to these characteristics, and other genomic characteristics such as polyploidy and DNA contamination are also likely to be estimated. The k-mer frequency curves also provide comprehensive information about the sequencing characteristics, such as error rate and degree of coverage bias. The proper processing of these issues will be important for the accurate estimation of genomic characteristics.

In addition to these theoretical models, we provide a set of programs for practical applications, which can be freely accessed from the BGI ftp site: ftp://ftp.genomics.org.cn/pub/gce.

K-mer frequency analysis has a significant accuracy advantage over traditional experimental technologies. In particular, genome projects utilizing next generation sequencing technologies often produce a higher coverage (>30X) of data, which makes the k-mer analysis more accurate compared to the low-coverage (<10X) required by traditional Sanger sequencing projects. We introduced the new k-mer individuals curve and float precision estimation method, which has the potential to increase the estimation accuracy by one or two magnitudes compared to the widely used rough integer precision estimation method used by many genome projects. On simulated sequencing data, our model achieved a very high estimation accuracy, and we analyzed how data coverage, repeat content, heterozygous rate, sequencing error rate and coverage bias degree can affect the estimation accuracy.

For real sequencing data, our model can deal with raw data, or one can perform low-quality filtering or error correction before k-mer counting. Reducing the errors in raw reads will decrease the computer memory consumption and make genomic characteristic estimation easier; however, the error processing methods may have some system bias and thus affect the accuracy of estimation. Our results show that both standard and continuous models can be applied to estimate the genome size and $a_i$ values when coverage bias exists. Although the estimation accuracy on real sequencing data is lower than that on simulated data, this degree of accuracy is good enough for many applications, such as helping to determine the sequencing strategy and guiding the development of assembly algorithms. To obtain accurate estimation for complex genomes from real sequencing data is still a great challenge, and future work should be focused on deeper understanding of those sequencing characteristics and developing advanced methods to process them in order to improve the estimation accuracy of genomic characteristics.

## List of abbreviations

***NGS***, next generation sequencing.

***GSP***, the name of Shan's program, genome size prediction

***GCE***, the name of our program, genome characteristics estimation.

***DBG***, de bruijn graph, most NGS assemblers are designed based on this algorithm.

***K-mer genomic frequency***, the appearance times of a specific k-mer in the genome.

***K-mer coverage depth***, the times that a specific k-mer being sequenced.

***K-mer species curve***, the traditional well-known k-mer coverage depth distribution curve showing the ratio of k-mer species classified by each coverage depth.

***K-mer individuals curve***, the novel k-mer coverage depth distribution curve introduced in this paper, showing the ratio of k-mer individuals classified by each coverage depth.

***Integer precision estimation***, obtain the $c$ value by directly observing the peak depth on the k-mer coverage depth distribution curves, which is often used as a rough estimation method.

***Float precision estimation***, our newly developed method to calculate $c$ value, using the relationships between the neighboring points in the k-mer coverage depth distribution curves, which has been adopted by the basic model, standard model and heterozygous model in this paper.

***Basic model***, the probability model of k-mer species and individuals curves for the simplest "ideal" genomes, which is equivalent to a random sequence.

***Standard model***, the probability model of k-mer species and individuals curves for the genomes with repeats.

***Heterozygous model***, the probability model of k-mer species and individuals curves for the genomes with repeats and heterozygosity.

***Continuous model***, the probability model for genome sequencing with coverage bias.

***Real curve***, the k-mer species or individuals curves plotted using real sequencing data.

***Theoretic curve***, the k-mer species or individuals curves plotted with theoretic values calculated by the formulas in each models with given reference $c$ and $a_i$ values.

***Estimated curve***, the k-mer species or individuals curves plotted with theoretic values calculated by the formulas in each models with given estimated $c$ and $a_i$ values.

## Competing interests

We have no competing interests to declare.

## Authors' contributions

WF and BL designed the study and drafted the manuscript. BL, YS and JY performed the statistical analysis and wrote the programs. XH, YT, HZ, NL, ZL, YC, JL, DM and SL participated in discussion, confirming the results and revising the manuscript. All authors read and approved the final manuscript.

## Acknowledgements

We are grateful to Junjie Qin, Junhua Li, Dongfang Li, Guojie Zhang, Cai Li, Shifeng Cheng for providing the sequencing data. We thank the members in the Science and Technology Department of BGI-SZ for various helpful discussions. We also thank Scott Edmunds, Yuexi tan, Jinsen Li, and Sijia Lu for polishing the English language.

Funding: This work was supported by the Basic Research Program Supported by Shenzhen City (grants JC2010526019), and the Key Laboratory Project Supported by Shenzhen City (grants

CXB200903110066A; CXB201108250096A).

# References

1. Pettersson E, Lundeberg J, Ahmadian A: **Generations of sequencing technologies**. *Genomics* 2009, **93**(2):105-111.
2. Li R, Fan W, Tian G, Zhu H, He L, Cai J, Huang Q, Cai Q, Li B, Bai Y *et al*: **The sequence and de novo assembly of the giant panda genome**. *Nature*, **463**(7279):311-317.
3. Xu X, Pan S, Cheng S, Zhang B, Mu D, Ni P, Zhang G, Yang S, Li R, Wang J *et al*: **Genome sequence and analysis of the tuber crop potato**. *Nature*, **475**(7355):189-195.
4. Jurka J, Kapitonov VV, Kohany O, Jurka MV: **Repetitive sequences in complex genomes: structure and evolution**. *Annual review of genomics and human genetics* 2007, **8**:241-259.
5. Otto SP: **The evolutionary consequences of polyploidy**. *Cell* 2007, **131**(3):452-462.
6. Barriere A, Yang SP, Pekarek E, Thomas CG, Haag ES, Ruvinsky I: **Detecting heterozygosity in shotgun genome assemblies: Lessons from obligately outcrossing nematodes**. *Genome research* 2009, **19**(3):470-480.
7. Alkan C, Sajjadian S, Eichler EE: **Limitations of next-generation genome sequence assembly**. *Nature methods*, **8**(1):61-65.
8. Birney E: **Assemblies: the good, the bad, the ugly**. *Nature methods*, **8**(1):59-60.
9. Dolezel J, Greilhuber J, Suda J: **Estimation of nuclear DNA content in plants using flow cytometry**. *Nature protocols* 2007, **2**(9):2233-2244.
10. Waring M, Britten RJ: **Nucleotide sequence repetition: a rapidly reassociating fraction of mouse DNA**. *Science (New York, NY* 1966, **154**(750):791-794.
11. Dewoody YD, Dewoody JA: **On the estimation of genome-wide heterozygosity using molecular markers**. *The Journal of heredity* 2005, **96**(2):85-88.
12. Gresham D, Curry B, Ward A, Gordon DB, Brizuela L, Kruglyak L, Botstein D: **Optimized detection of sequence variation in heterozygous genomes using DNA microarrays with isothermal-melting probes**. *Proceedings of the National Academy of Sciences of the United States of America* 2010, **107**(4):1482-1487.
13. Huang S, Li R, Zhang Z, Li L, Gu X, Fan W, Lucas WJ, Wang X, Xie B, Ni P *et al*: **The genome of the cucumber, Cucumis sativus L**. *Nature genetics* 2009, **41**(12):1275-1281.
14. Nygaard S, Zhang G, Schiott M, Li C, Wurm Y, Hu H, Zhou J, Ji L, Qiu F, Rasmussen M *et al*: **The genome of the leaf-cutting ant Acromyrmex echinatior suggests key adaptations to advanced social life and fungus farming**. *Genome research*.
15. Li X, Waterman MS: **Estimating the repeat structure and length of DNA sequences using L-tuples**. *Genome research* 2003, **13**(8):1916-1922.
16. GAO SHAN W-MZ: **An $\ell$-mer component distribution for genome size esimation.** *Sourceforge* 2009, http://sourceforge.net/projects/gsizepred/.
17. Myers EW, Sutton GG, Delcher AL, Dew IM, Fasulo DP, Flanigan MJ, Kravitz SA, Mobarry CM, Reinert KH, Remington KA *et al*: **A whole-genome assembly of Drosophila**. *Science (New York, NY* 2000, **287**(5461):2196-2204.
18. Kurtz S, Narechania A, Stein JC, Ware D: **A new method to compute K-mer frequencies and its application to annotate large repetitive plant genomes**. *BMC genomics* 2008, **9**:517.
19. Marcais G, Kingsford C: **A fast, lock-free approach for efficient parallel counting of occurrences of k-mers**. *Bioinformatics (Oxford, England)* 2011, **27**(6):764-770.
20. Staden R: **A strategy of DNA sequencing employing computer programs**. *Nucleic acids research* 1979, **6**(7):2601-2610.
21. Arratia R, Martin D, Reinert G, Waterman MS: **Poisson process approximation for sequence repeats, and sequencing by hybridization**. *J Comput Biol* 1996, **3**(3):425-463.
22. Wicker T, Sabot F, Hua-Van A, Bennetzen JL, Capy P, Chalhoub B, Flavell A, Leroy P, Morgante M, Panaud O *et al*: **A unified classification system for eukaryotic transposable elements**. *Nature reviews* 2007, **8**(12):973-982.
23. Miller JR, Koren S, Sutton G: **Assembly algorithms for next-generation sequencing data**. *Genomics*, **95**(6):315-327.

24. Bentley DR, Balasubramanian S, Swerdlow HP, Smith GP, Milton J, Brown CG, Hall KP, Evers DJ, Barnes CL, Bignell HR *et al*: **Accurate whole human genome sequencing using reversible terminator chemistry**. *Nature* 2008, **456**(7218):53-59.
25. Nakamura K, Oshima T, Morimoto T, Ikeda S, Yoshikawa H, Shiwa Y, Ishikawa S, Linak MC, Hirai A, Takahashi H *et al*: **Sequence-specific error profile of Illumina sequencers**. *Nucleic acids research*.
26. Dohm JC, Lottaz C, Borodina T, Himmelbauer H: **Substantial biases in ultra-short read data sets from high-throughput DNA sequencing**. *Nucleic acids research* 2008, **36**(16):e105.
27. Li DX, F; Zhao, M; Chen, W; Cao, S; Xu, R; Wang, G; Wang, J; Zhang, Z; Li, Y; Cui, C; Chang, C; Cui, C; Luo, Y; Qin, J; Li, S; Li, J; Peng, Y; Pu, F; Sun, Y; Chen, Y; Zong, Y; Ma, X; Yang, X; Cen, Z; Song, Y; Zhao, X; Chen, F; Yin, X; Rohde, H; Liang, Y; Li, Y and the Escherichia coli O104:H4 TY-2482 isolate genome sequencing consortium: **Genomic data from Escherichia coli O104:H4 isolate TY-2482**. *BGI Shenzhen* 2011, http://dx.doi.org/10.5524/100001.
28. Li R, Zhu H, Ruan J, Qian W, Fang X, Shi Z, Li Y, Li S, Shan G, Kristiansen K *et al*: **De novo assembly of human genomes with massively parallel short read sequencing**. *Genome research*, **20**(2):265-272.
29. Wang J, Wang W, Li R, Li Y, Tian G, Goodman L, Fan W, Zhang J, Li J, Guo Y *et al*: **The diploid genome sequence of an Asian individual**. *Nature* 2008, **456**(7218):60-65.
30. Kelley DR, Schatz MC, Salzberg SL: **Quake: quality-aware detection and correction of sequencing errors**. *Genome biology*, **11**(11):R116.

# Figures

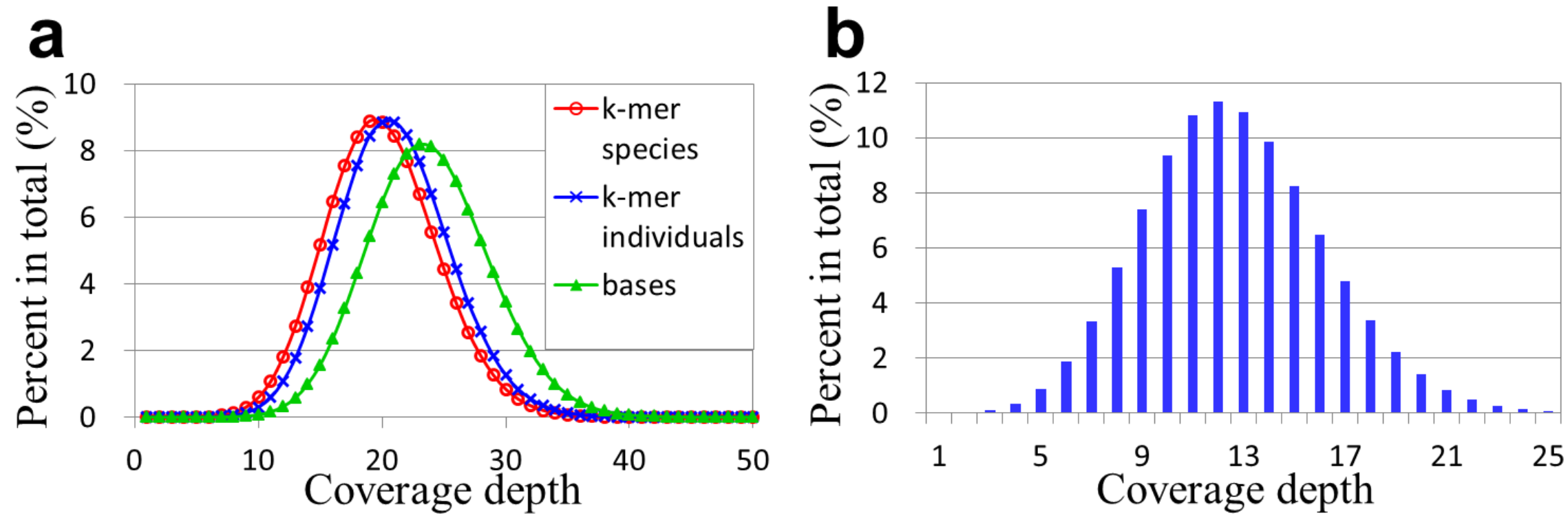

**Figure 1. Illustration of the basic distribution model by "ideal" genome and sequencing data.**

**a.** Distributions of base and k-mer coverage depth plotted with 23.8X simulated "ideal" sequencing data generated from the 10-Mb "ideal" reference genome. The read length ($L$) and k-mer size ($K$) used here is 100 and 17 respectively, so the expected k-mer coverage depth is 20. The depth information in the bases curve is obtained by mapping the reads onto the reference genome and counting the coverage depth at each genomic location, whist those in the k-mer species and individuals curves are obtained by counting the k-mer frequency using sequencing data, all of which are consistent with the theoretic curves constructed by $c$ and reference $a_i$ values (not shown). Note that the number of k-mer individuals is the product of k-mer species number and corresponding depth value. **b.** The histogram of Poisson distribution with expected coverage depth of 12.6, but the observed peak depth value here is 12, which has roughly 5% difference from the expected value.

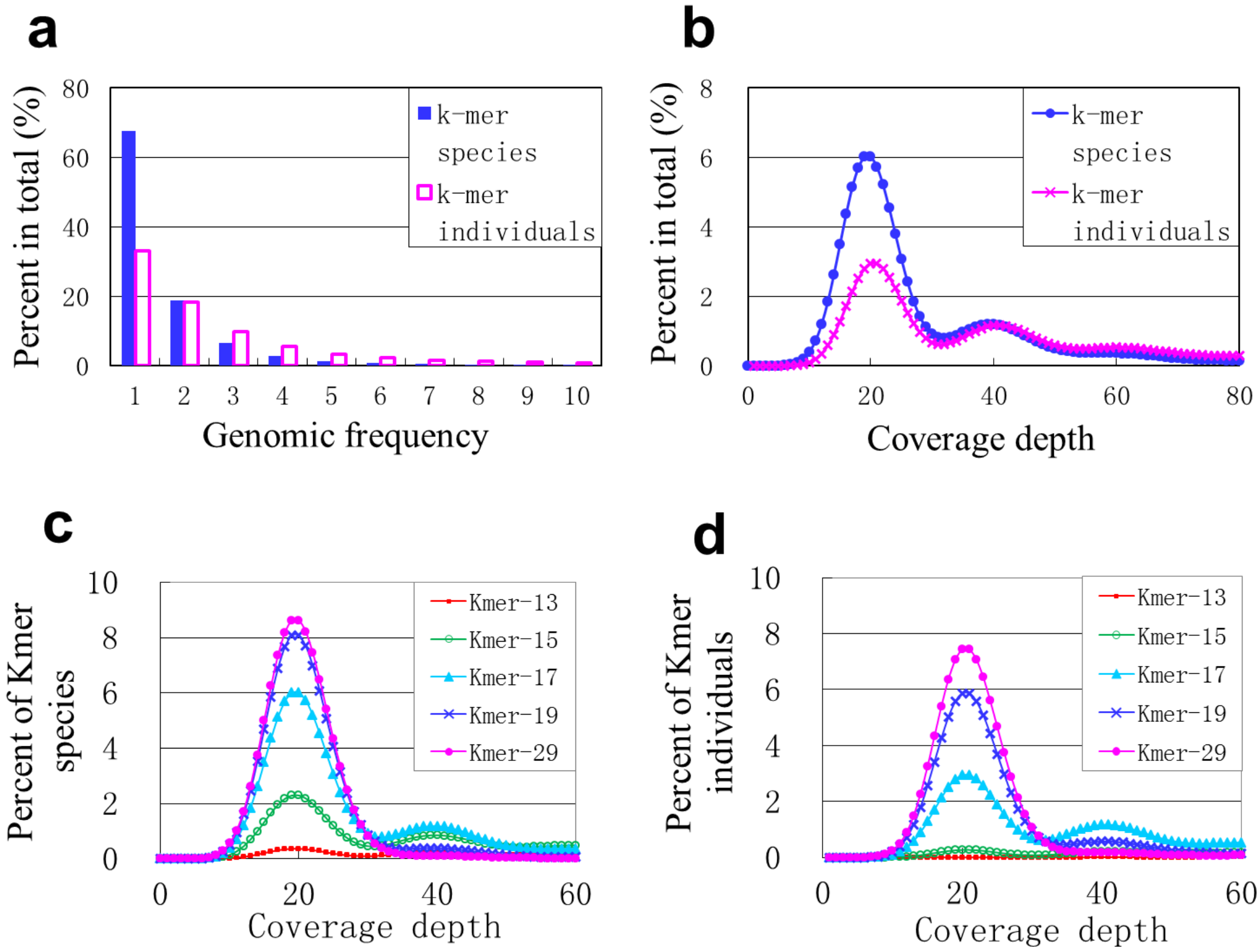

**Figure 2. Illustration of the standard model by repetitive genome (human) and "ideal" sequencing data.**

**a.** Distribution of genomic frequency for k-mers ($K$=17) in the human reference genome, the ratio of k-mer species ($a_i$) and individuals ($b_i$) were shown respectively. **b.** Distribution of coverage depth for k-mers ($K$=17) from the "ideal" sequencing data simulated based on the human reference genome ($c_{k\text{-}mer}$=20). The ratio of k-mer species and individuals were plotted respectively, both of which are well consistent with the theoretic curves constructed by the compound models with $c$ and reference $a_i$ values (not shown). **c**. The k-mer species curves for the human reference genome with "ideal" sequencing data ($c_{k\text{-}mer}$=20), with the k-mer sizes vary from 13 to 29. **d**. The k-mer individuals curves with the same conditions. Note that the k-mer sizes smaller than 17 are non-proper for the human genome, at which cases the curves are nearly flat and the major peak is not very dominant. In contrast, when using k-mer size up to 29, it seems that no repeat peaks can be obviously observed.

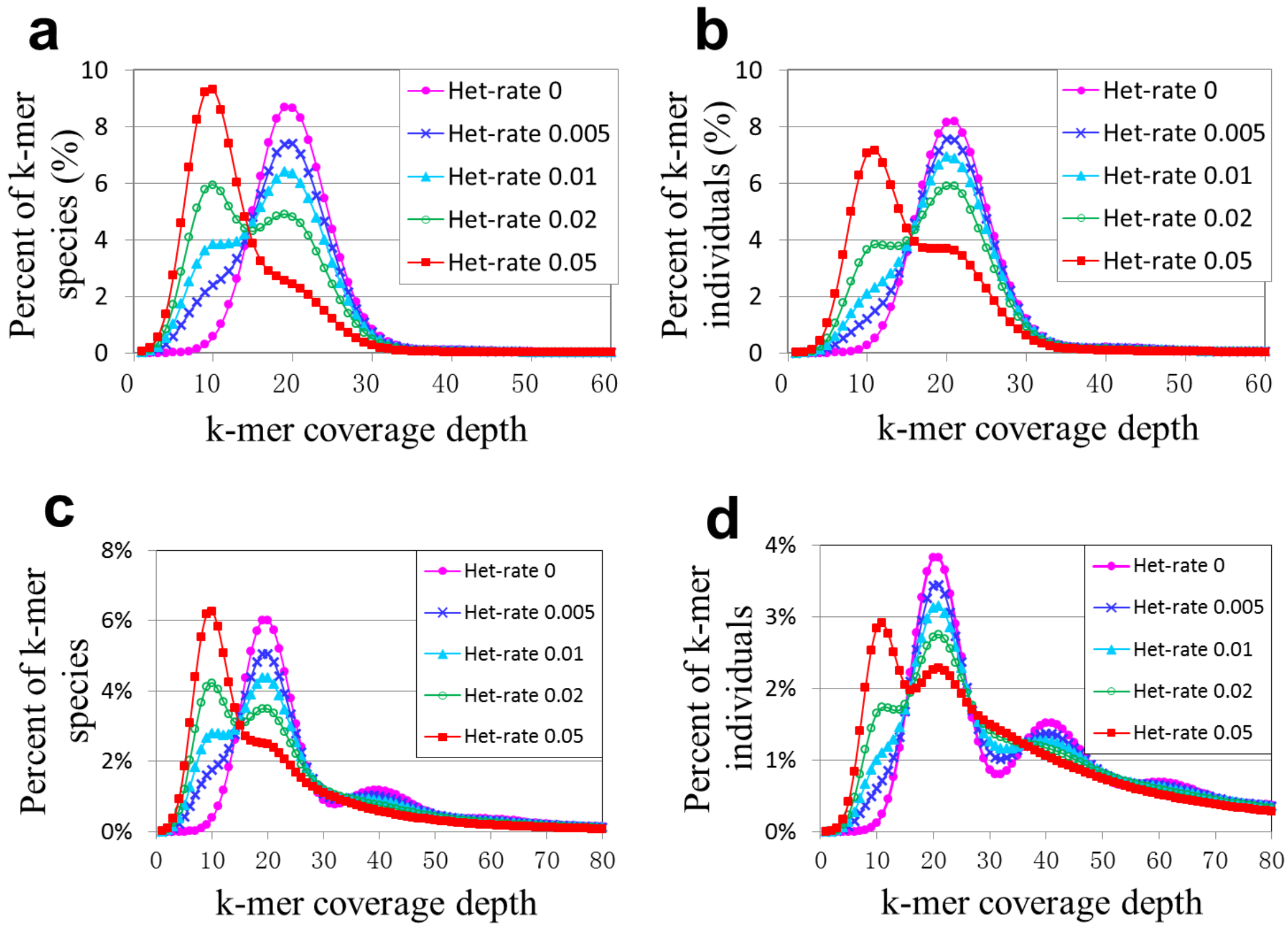

**Figure 3. Illustration of the heterozygous model using "ideal" sequencing data.**

**a** and **b** shows the k-mer ($K$=17) species and individuals curves for the simplest "ideal" genome, using simulated reads ($c_{k\text{-}mer}$=20) with various heterozygous rate ranging from zero to 5%. **c** and **d** shows the k-mer ($K$=17) species and individuals curves for the repetitive genome (human), using simulated reads ($c_{k\text{-}mer}$=20) with various heterozygous rate ranging from zero to 5%.

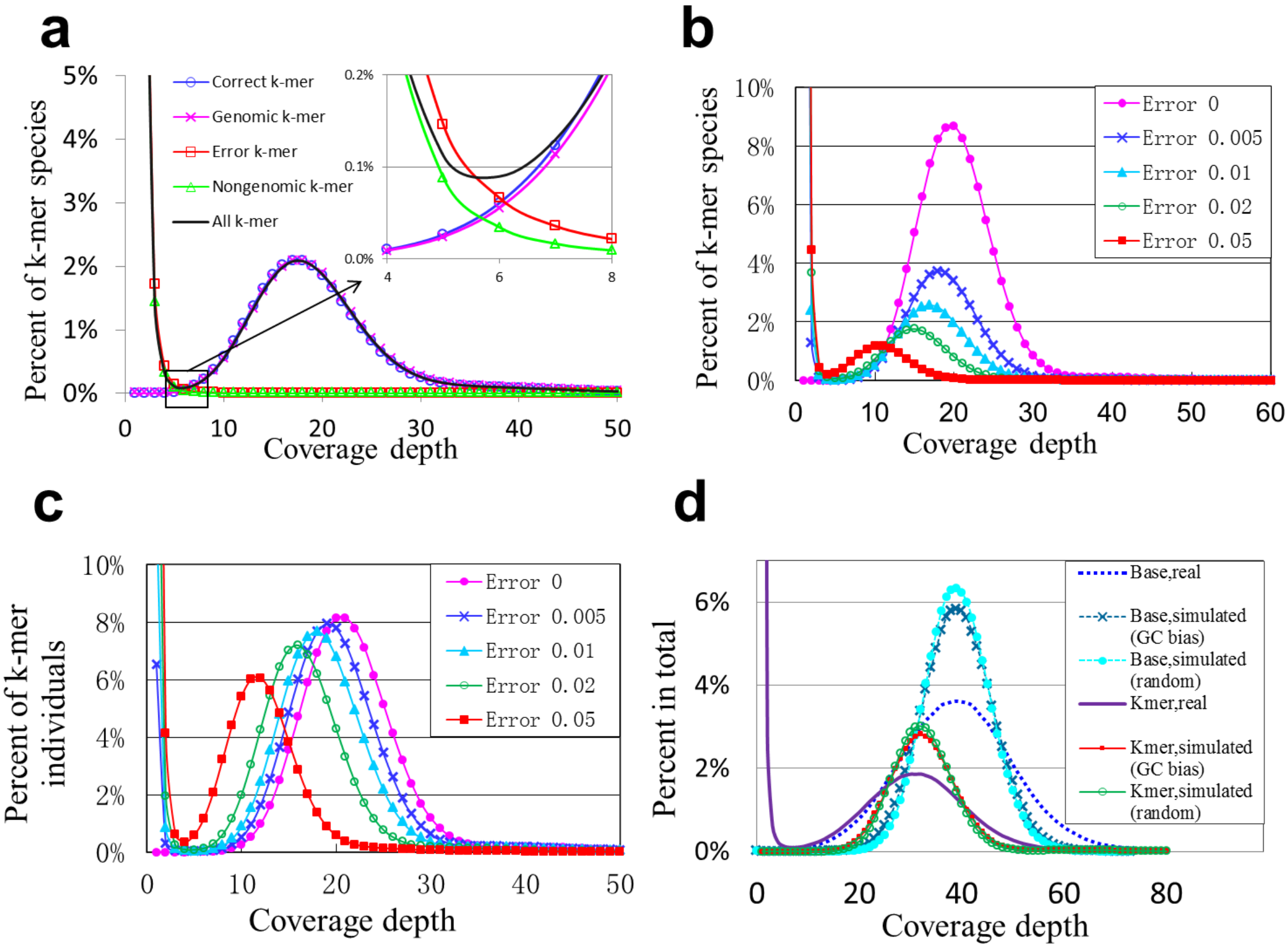

**Figure 4. The influence of sequencing error on the k-mer species curve.**

**a.** The influence of sequencing error on k-mer species curve. The k-mer size was set to be 17 for simulated reads with 1% error rate from the non-heterozygous Arabidopsis genome. Five types of k-mer species curves were plotted for all k-mers, genomic k-mers, non-genomic k-mers, correct k-mers and erroneous k-mers respectively. **b** and **c** shows the k-mer ($K$=17) species curves and individuals curves using the simulate reads ($c_{k\text{-}mer}$=20) from simplest "ideal" genome with various rate of sequencing errors ranging from zero to 5%. **d**. The influence of sequencing coverage bias on k-mer species curve ($K$=17) shown by the Germany Ecoli data. The coverage distribution for both bases and k-mers were plotted, each with 3 curves represented for real sequencing data, simulated sequencing data with GC-bias, and randomly simulated sequencing data respectively. Note that we firstly mapped the real sequencing reads onto the assembled Germany Ecoli reference genome and calculated the relationship between GC content and coverage depth using 100-bp windows, and then the resulting model was used to simulate sequencing data with GC-related coverage bias, which was used to plot the GC-bias related curves.

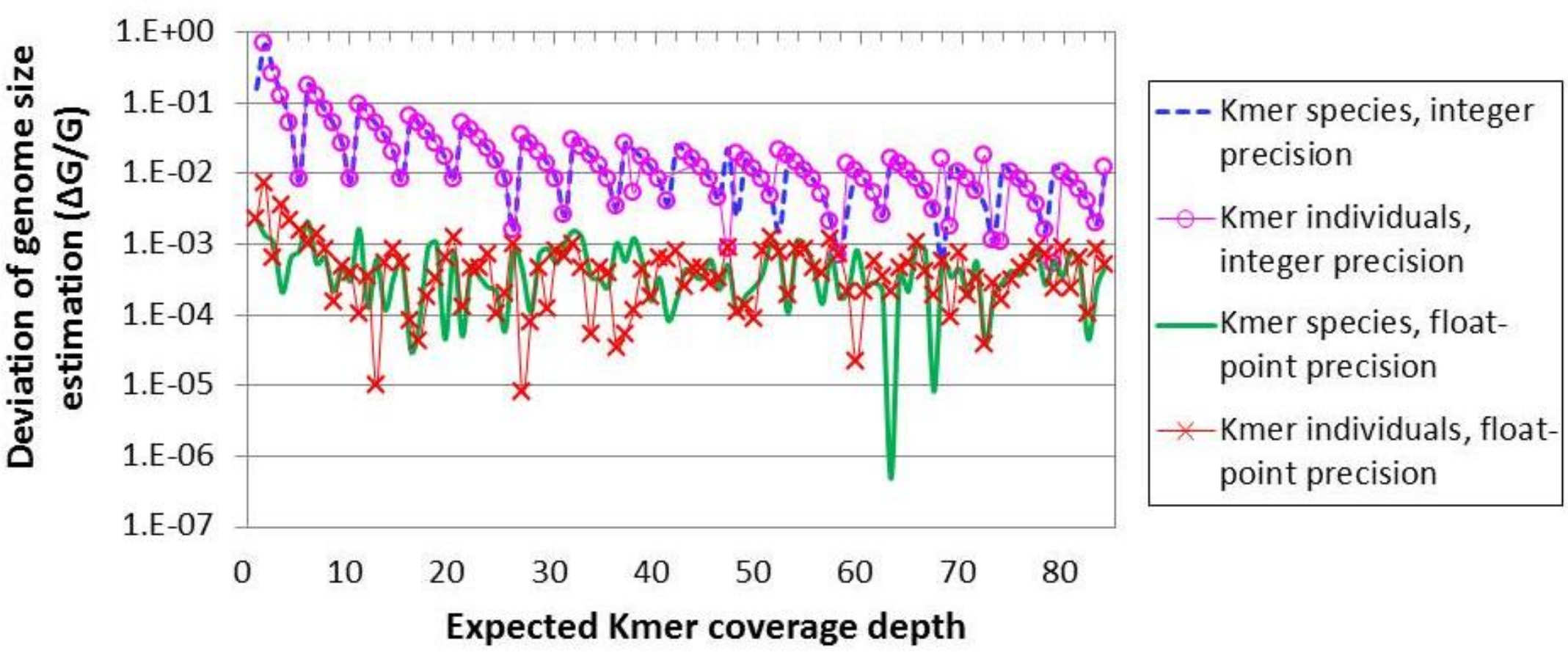

**Figure 5**. **Deviation of genome size estimation (ΔG/G) with various expected k-mer coverage depth using "ideal" reference genome and sequencing data.**

The sequencing depth ($c_{base}$) is ranging from 1 to 100 with step 1, and so the $c_{k\text{-}mer}$ ($0.84{*}c_{base}$) is ranging from 0.84 to 84. The ratio of genome size deviation is represented by logarithmic scale. Four curves were plotted to show results from integer and float precision estimation based on the k-mer species and individuals curve respectively.

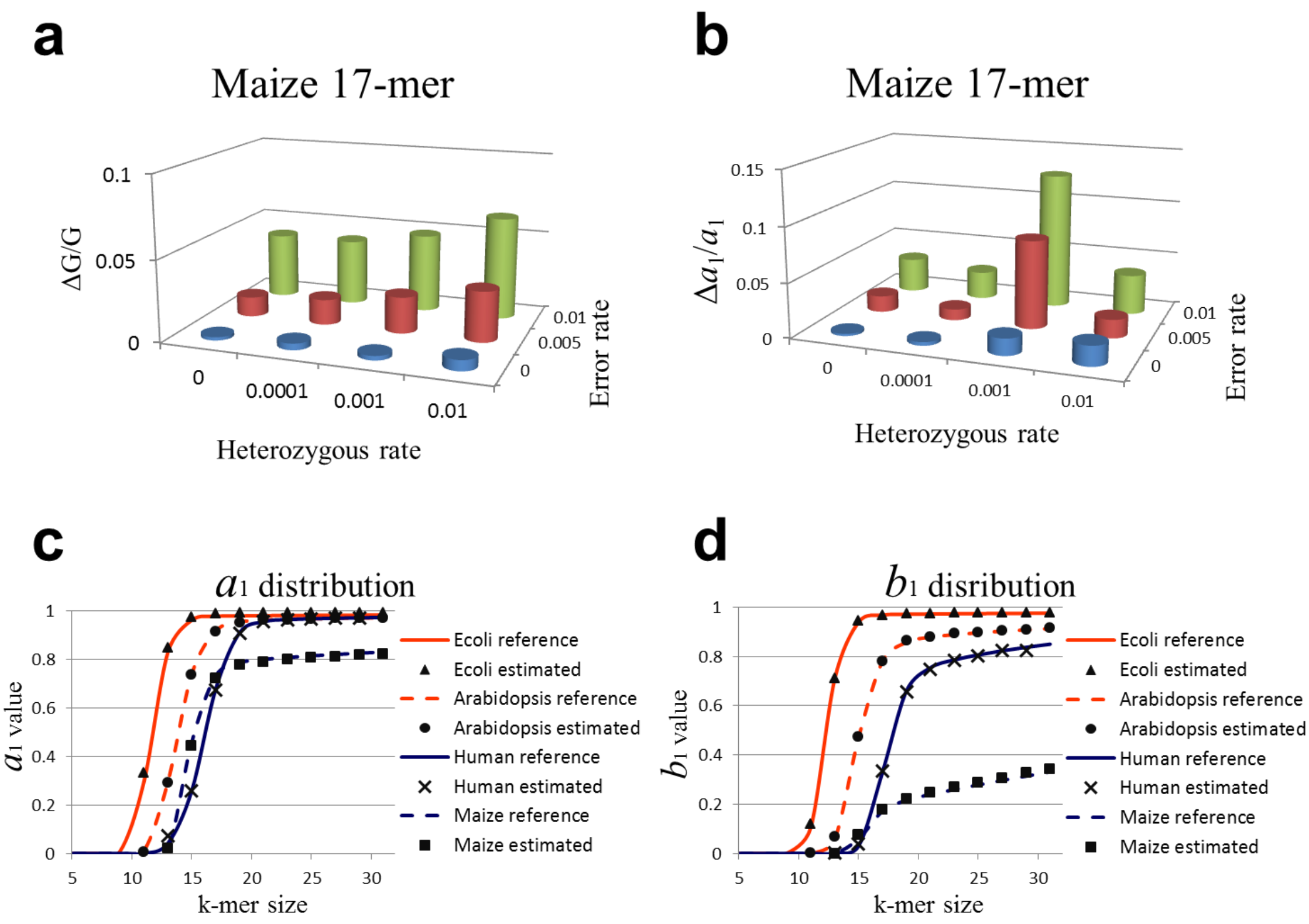

**Figure 6. The estimation accuracy with simulated data.**

**a and b.** Distribution of $\Delta G/G$ and $\Delta a_1/a_1$ values with k-mer size 17 for maize genome, which has the most repeat content among the 4 reference species. The $\Delta G/G$ or and $\Delta a_1/a_1$ values under each combination of heterozygous rate and error rate are shown respectively. The Figures for the other 3 species can be found in Figure S2 and Figure S3. **c and d.** The distribution of $a_1$ and $b_1$ values with various k-mer sizes for the 4 species. The curves were plotted using values calculated by the reference genome sequences, whist the highlighted points were calculated from the simulated sequencing data with 0.5% error rate on non-heterozygous genome.

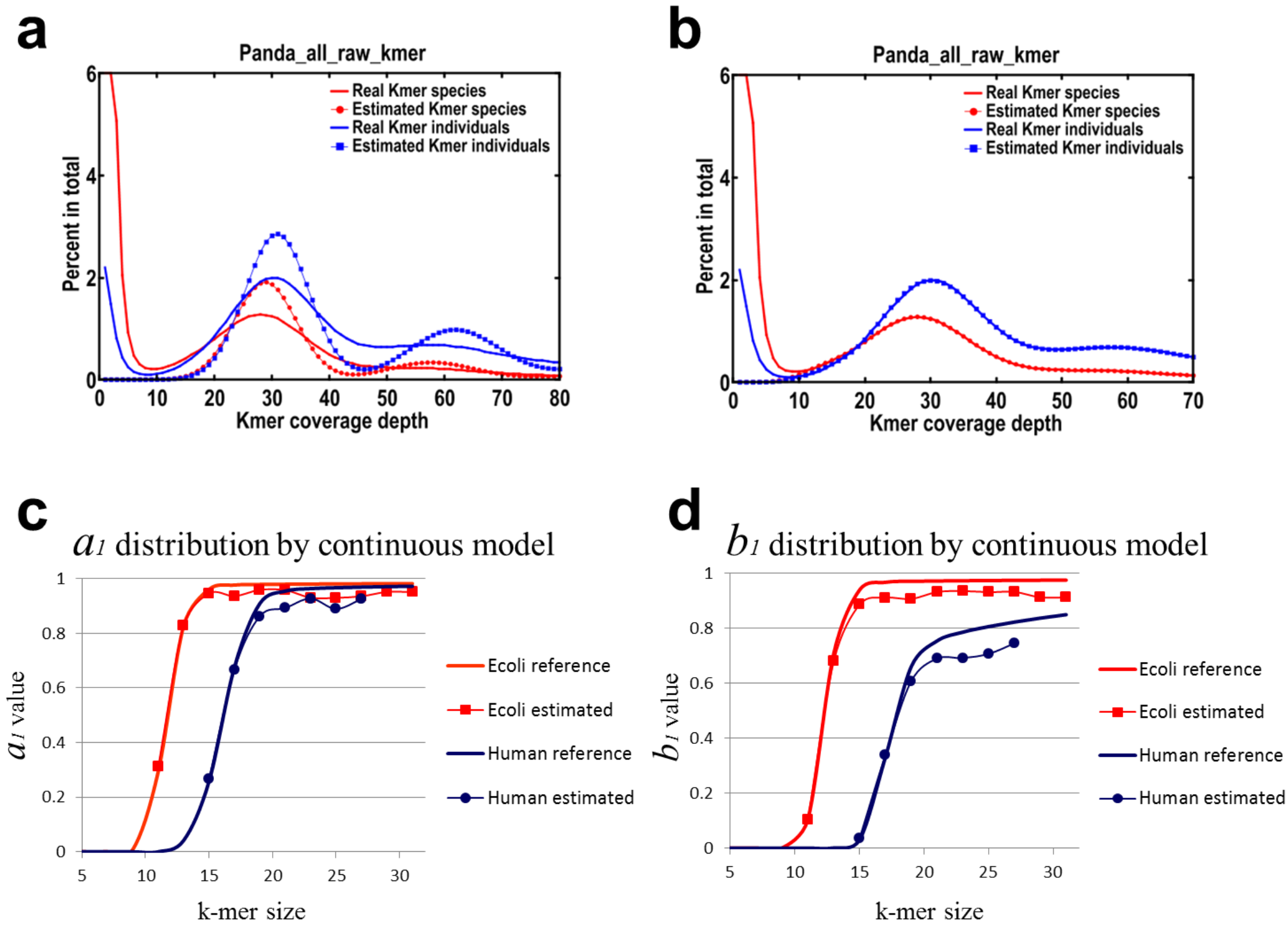

**Figure 7. The estimation accuracy with real sequencing data.**

**a and b**. The real k-mer species and individuals curves of panda data, as well as the estimated curves using the standard model and continuous model, respectively. **c and d.** The distribution of $a_1$ and $b_1$ values with various k-mer sizes for Germany Ecoli and Yanhuang (human). As these two species have relatively fine assembled reference genomes, so the estimated values from sequencing data can be compared to that calculated from genomic sequences. The estimated results from the standard model and continuous model are similar, and so only results from the continuous model are shown.

# Tables

**Table 1. Genome sizes and reference $a_i$ values for the 4 reference species.**

| Species | Assembly version | Genome size | $a_1$ | $a_2$ | $a_3$ | $a_4$ | $a_5$ |
|---|---|---|---|---|---|---|---|
| Ecoli | NC_000913 | 4,639,675 | 99.06% | 0.50% | 0.17% | 0.05% | 0.04% |
| Arabidopsis | TIGR Release 5.0 | 118,997,677 | 91.66% | 5.98% | 1.14% | 0.43% | 0.23% |

| Human | NCBI36 | 2,832,359,852 | 67.62% | 18.73% | 6.60% | 2.79% | 1.38% |
|---|---|---|---|---|---|---|---|
| Maize | release-4a.53 | 2,033,474,564 | 73.17% | 12.59% | 4.46% | 2.38% | 1.47% |

Note: The reference genome sequences were downloaded from public databases: Maize from ftp://ftp.maizesequence.org/pub/maize/, and the others from http://www.ncbi.nlm.nih.gov/genome/. The gap sequences were removed from the reference sequences before doing any further analysis, and this table shows the first 5 $a_i$ values of 17-mer with no heterozygosity.

**Table 2. Reference and estimated genome sizes for real sequencing data.**

a. The background of species genome used in real sequencing data.

| Species | genome size in paper | assembly length | $a_1$ of assembly seq | assembly cvg ratio |
|---|---|---|---|---|
| Ant | 313,000,000 | 299,573,819 | 87.05% | 95.71% |
| E.coli | | 5,444,474 | 97.71% | |
| Human | 3,000,000,000 | 2,832,359,852 | 67.62% | 94.41% |
| Panda | 2,400,000,000 | 2,299,498,912 | 72.01% | 95.81% |
| Potato | 844,000,000 | 705,875,680 | 77.31% | 83.63% |

Note: The $a_1$ value is calculated with k-mer size 17 from the assembled genome sequences for each species, which is not complete and usually missing repeat sequences.

b. Estimated genome sizes with different data sets and models.

| Methods \ Species | | Ecoli | Ant | Potato | Panda | Yanhuang |
|---|---|---|---|---|---|---|
| Reported in paper | | 5,444,000 | 313,000,000 | 844,000,000 | 2,400,000,000 | 3,000,000,000 |
| Raw data | Rough | 5,549,866 | 344,523,816 | 861,678,589 | 2,563,568,959 | 3,030,022,547 |
| | Standard | 5,270,968 | 311,699,465 | 791,473,035 | 2,347,329,461 | 2,909,815,725 |
| | Continuous | 5,304,537 | 299,968,437 | 809,264,229 | 2,396,293,201 | 2,983,982,946 |
| With error correction | Rough | 5,600,499 | 311,867,315 | 793,498,707 | 2,409,346,995 | 2,946,704,974 |
| | Standard | 5,428,390 | 303,181,843 | 775,067,312 | 2,332,652,503 | 2,870,287,246 |
| | Continuous | 5,204,046 | 309,661,529 | 786,567,649 | 2,370,340,602 | 2,946,741,792 |
| Filter low-quality k-mers | Rough | 5,848,517 | 343,151,351 | 831,580,810 | 2,571,111,447 | 3,051,272,564 |
| | Standard | 5,601,505 | 325,397,589 | 823,523,416 | 2,415,518,152 | 2,918,462,646 |
| | Continuous | 5,400,141 | 311,771,512 | 814,396,835 | 2,326,443,641 | 2,850,470,598 |

Note: The previously reported genome size values for Ecoli and Yanhuang (human) are most reliable, which have nearly complete reference genomes; however, all the reported values are not totally accurate and should be doubted in the analysis. The assembled genome sequences as well as the raw sequencing data were downloaded from the URL listed in the each published genome paper.

**Table 3. Comparison of estimation accuracy between GCE and GSP.**

a. Use simulated data without sequencing errors.

| species | GCE | | GSP | |
|---|---|---|---|---|
| | $\Delta G/G$ | $\Delta a_1/a_1$ | $\Delta G/G$ | $\Delta a_1/a_1$ |
| Ecoli | 0.15% | 0.06% | 0.32% | 0.07% |
| Arabidopsis | 4.21% | 0.70% | 0.15% | 0.14% |
| Human | 38.31% | 26.59% | 0.14% | 0.24% |

| Maize | 74.50% | 29.15% | 0.21% | 0.20% |
|---|---|---|---|---|

b. Use simulated data with 1% error rate

| species | GCE | | GSP | |
|---|---|---|---|---|
| | $\Delta G/G$ | $\Delta a_1/a_1$ | $\Delta G/G$ | $\Delta a_1/a_1$ |
| Ecoli | 0.10% | 72.84% | 0.12% | 0.05% |
| Arabidopsis | 3.12% | 70.98% | 0.41% | 1.59% |
| Human | 37.77% | 48.98% | 0.57% | 1.13% |
| Maize | 71.89% | 70.21% | 4.07% | 3.19% |

Note: we simulated 25X reads with no heterozygous rate from the reference sequence of Arabidopsis, E.coli, Human and Maize. The k-mer size here is 17bp.we used parameter here for GSP estimation is –k 17 –r 100 –e 0, other parameters are set as default.

# Supplemental materials

## Supplementary figures

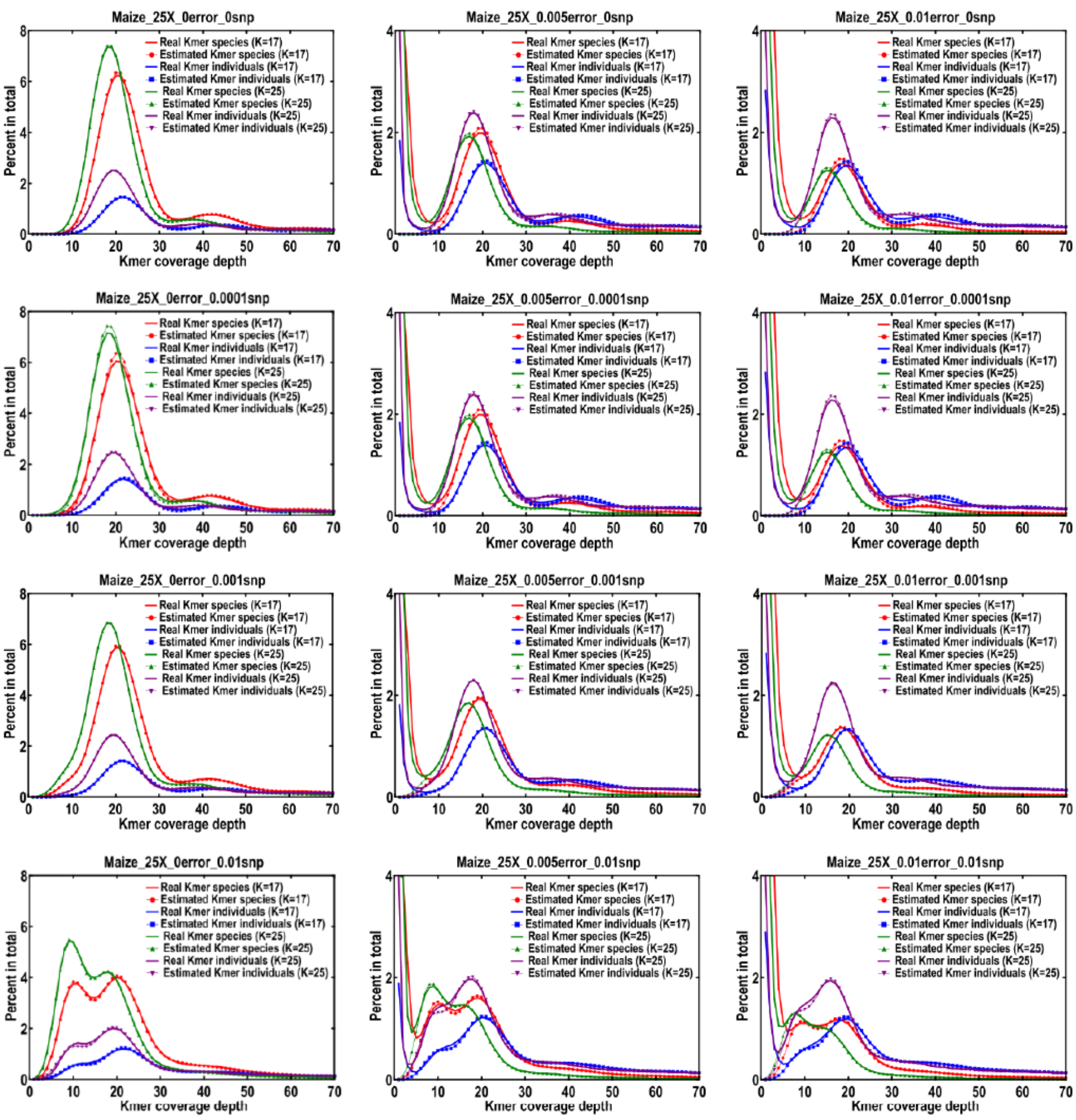

**Figure S1. K-mer species and individuals curves for the 48 data sets.**
This figure contains the k-mer species and individuals curves for all the testing data sets. On each sub figure, 8 curves were shown (17-mer and 25-mer , k-mer species and individuals , real and estimated). For most of the sub figures, the estimated curves are well consistent with the real curves.

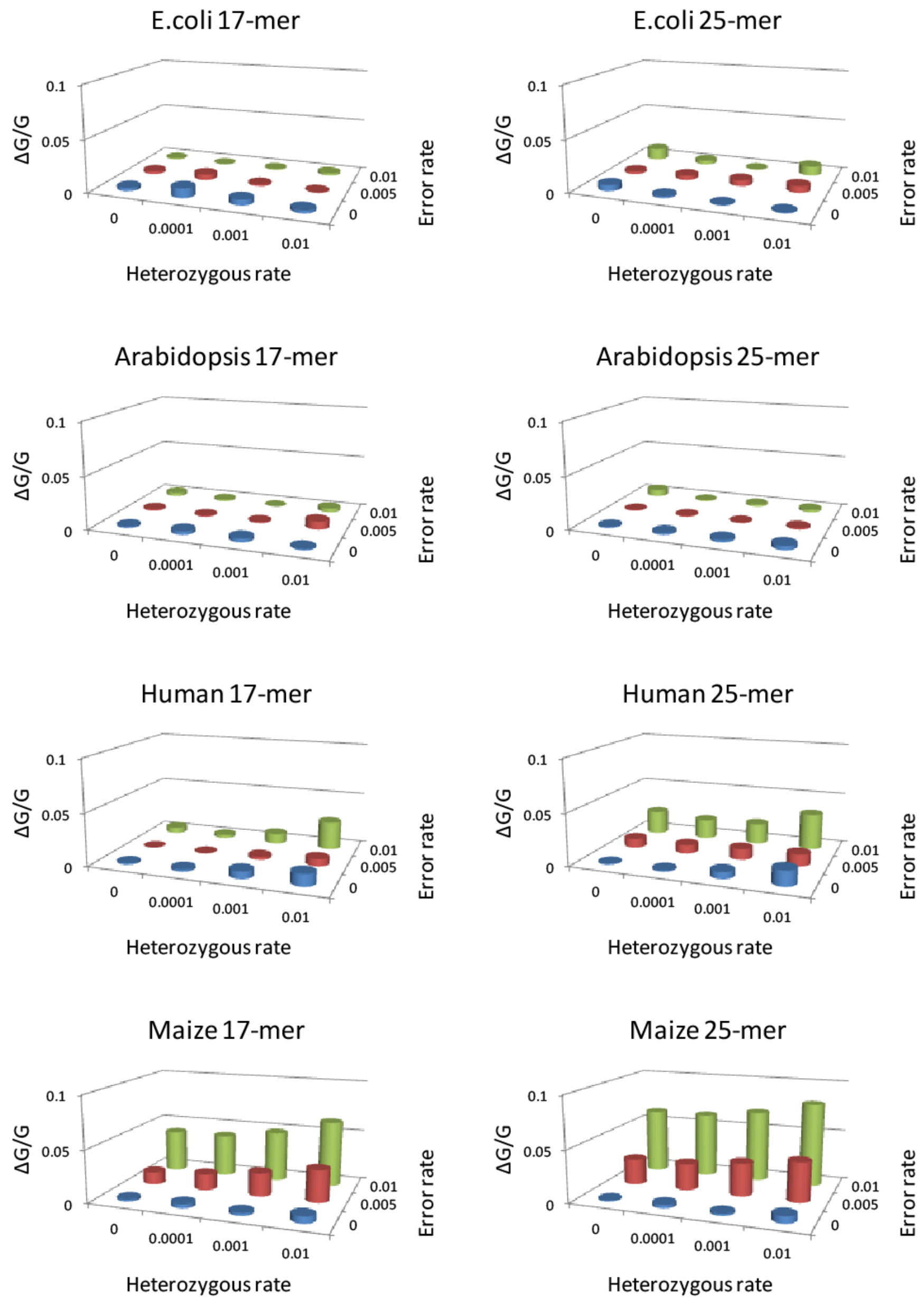

**Figure S2. Distribution of ΔG/G values for all the data sets.**
This figure contains the distribution of ΔG/G values with k-mer size 17 and 25 for 4 reference species using simulated data. In each sub figure, the ΔG/G values under each combination of heterozygous rate and error rate were shown.

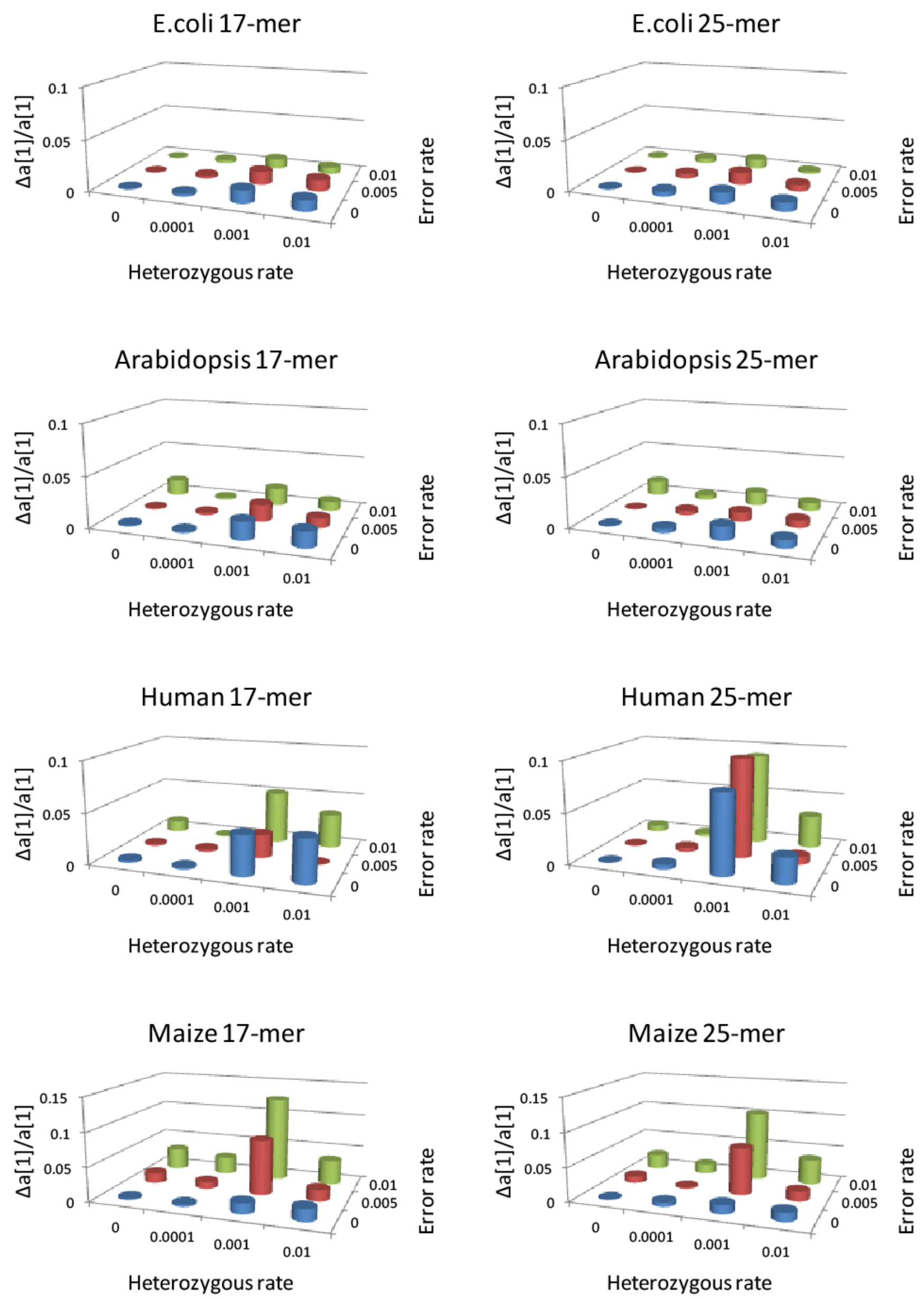

**Figure S3. Distribution of Δa1/a1 values for all the data sets.**
This figure contains the distribution of Δa1/a1 values with k-mer size 17 and 25 for 4 reference species using simulated data. In each figure, the Δa1/a1 values under each combination of heterozygous rate and error rate are shown.

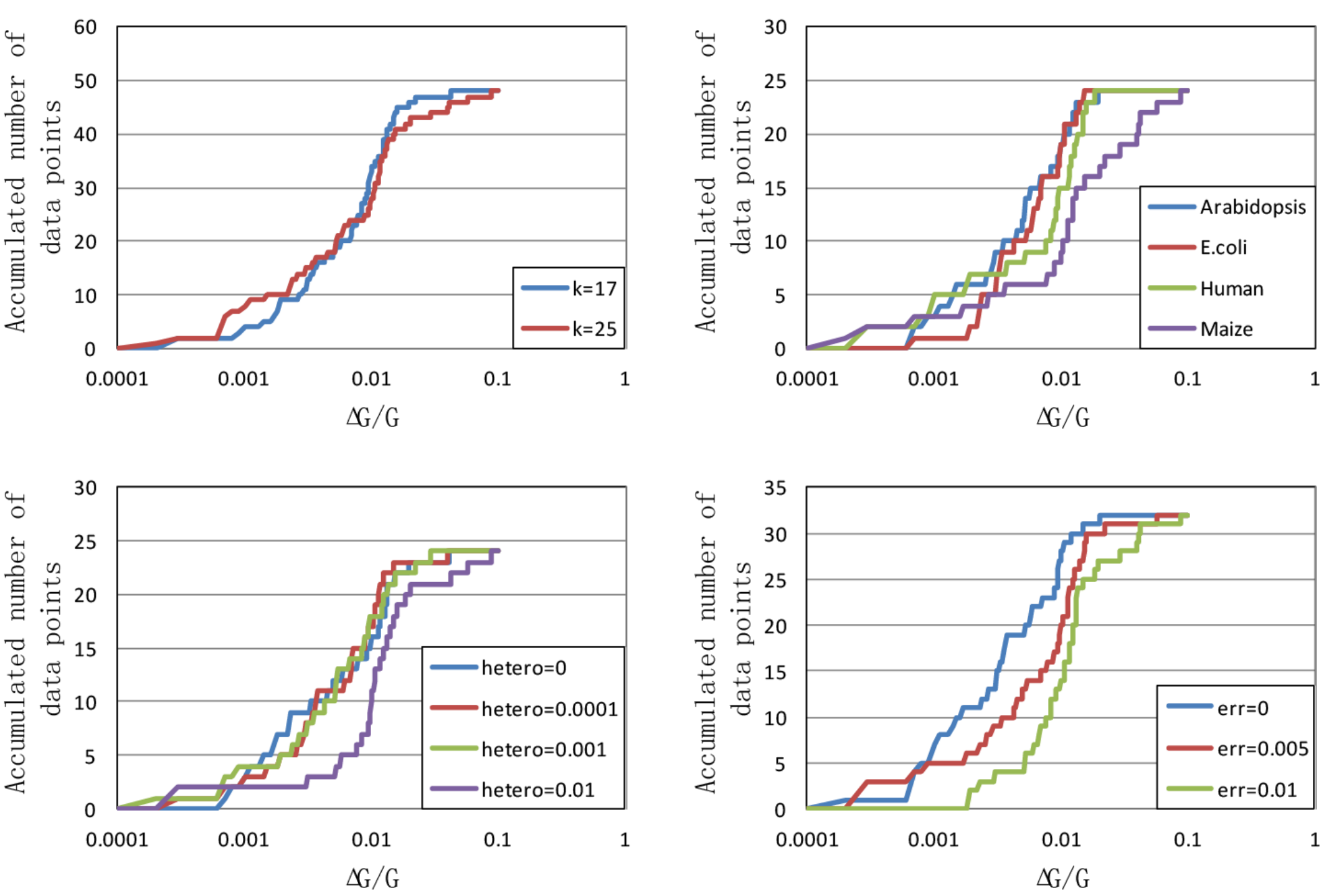

**Figure S4. Distribution of ΔG/G values separated by each affecting factor.**
This figure contains the distribution of ΔG/G values from the totally 96 analysis sets separated by k-mer sizes (a), heterozygous rate (b), error rate (c), and repeat content (d). Note that the X-axis is in logarithmic scale, whist the Y-axis means accumulated number of data points.

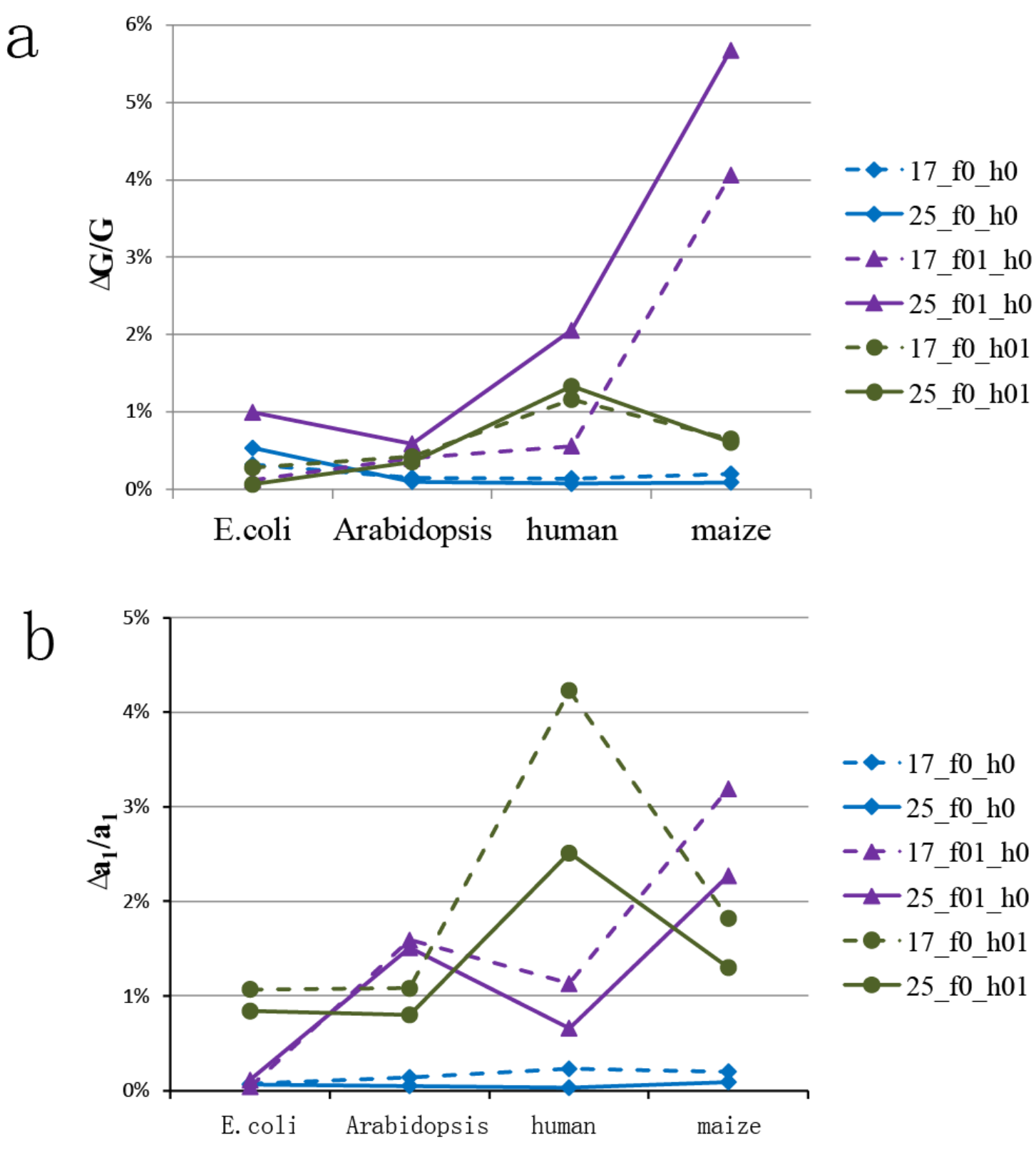

**Figure S5. the influence of k-mer size on estimation accuracy.**

This figure shows the influence of k-mer size on the accuracy of genome size estimation and ai estimation. All the estimation accuracy of 17-mer was shown with dash lines, and 25-mer were shown as full lines. The estimation accuracy from data with sequencing error (f01, 1%) was shown with triangle and the estimation accuracy from data with heterozygosis (h01, 1%) was shown with circle. The estimation result from data without sequencing error or heterozygous was shown with rhombus.

a. Using standard model :

b. Using continuous model:

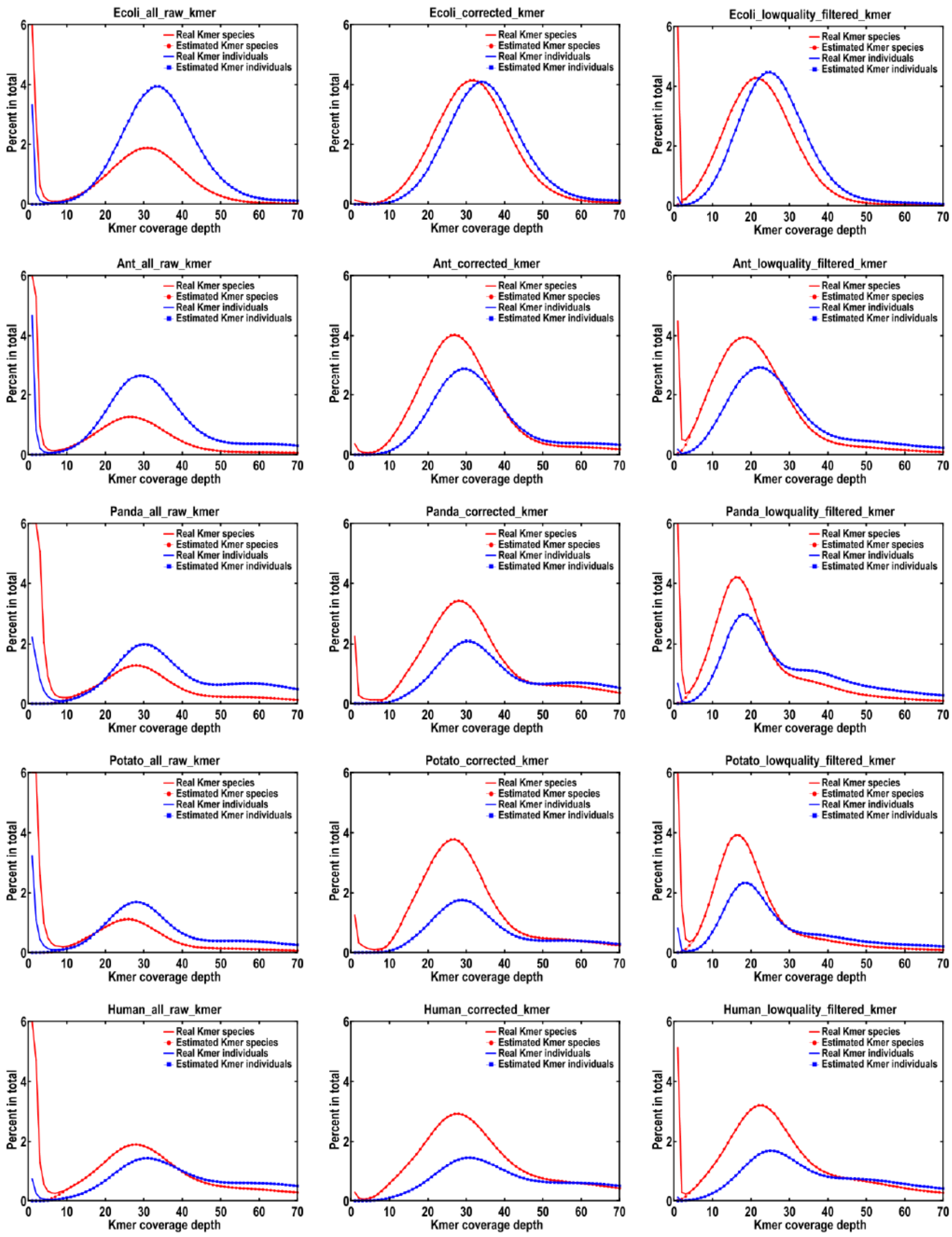

**Figure S6. The k-mer species and individuals curves for real data.**
This figure contains the k-mer species and individuals curve (K=17) for the 5 species with real sequencing data. Besides the real curves, the estimated curves by standard model and continuous model were also shown.

# Supplementary tables

**Table S1. All the reference $a_i$ values calculated from genomic sequences.**

| Species | SnpRate | KmerSize | a[0.5] | a[1] | a[1.5] | a[2] | a[2.5] | a[3] | a[3.5] | a[4] | a[4.5] | a[5] |
|---|---|---|---|---|---|---|---|---|---|---|---|---|
| Human | 0 | 17 | - | 0.6762 | - | 0.1873 | - | 0.066 | - | 0.0279 | - | 0.0138 |
| | | 25 | - | 0.9674 | - | 0.0185 | - | 0.0053 | - | 0.0025 | - | 0.0015 |
| | 0.0001 | 17 | 0.0027 | 0.6734 | 0.0013 | 0.186 | 0.0006 | 0.0654 | 0.0004 | 0.0276 | 0.0002 | 0.0136 |
| | | 25 | 0.0051 | 0.9623 | 0.0001 | 0.0183 | 0.0001 | 0.0053 | 0 | 0.0025 | 0 | 0.0015 |
| | 0.001 | 17 | 0.0258 | 0.6489 | 0.0122 | 0.1756 | 0.006 | 0.0604 | 0.0032 | 0.0249 | 0.0019 | 0.0121 |
| | | 25 | 0.0942 | 0.8251 | 0.0505 | 0.0147 | 0.0026 | 0.0037 | 0.0012 | 0.0016 | 0.0007 | 0.0009 |
| | 0.01 | 17 | 0.2014 | 0.4748 | 0.0776 | 0.1122 | 0.0316 | 0.0349 | 0.0141 | 0.0135 | 0.0071 | 0.0063 |
| | | 25 | 0.3646 | 0.607 | 0.0071 | 0.0093 | 0.0023 | 0.0024 | 0.0011 | 0.001 | 0.0007 | 0.0006 |
| Maize | 0 | 17 | - | 0.7317 | - | 0.1259 | - | 0.0446 | - | 0.0238 | - | 0.0147 |
| | | 25 | - | 0.8143 | - | 0.0883 | - | 0.0302 | - | 0.0164 | - | 0.0101 |
| | 0.0001 | 17 | 0.0047 | 0.7271 | 0.0012 | 0.1247 | 0.0006 | 0.044 | 0.0004 | 0.0234 | 0.0003 | 0.0144 |
| | | 25 | 0.0069 | 0.8078 | 0.001 | 0.0872 | 0.0005 | 0.0297 | 0.0003 | 0.0161 | 0.0002 | 0.0098 |
| | 0.001 | 17 | 0.0444 | 0.6889 | 0.0107 | 0.1149 | 0.005 | 0.0395 | 0.0031 | 0.0205 | 0.0022 | 0.0123 |
| | | 25 | 0.0645 | 0.7537 | 0.0082 | 0.0784 | 0.0039 | 0.0258 | 0.0025 | 0.0136 | 0.0018 | 0.0081 |
| | 0.01 | 17 | 0.2975 | 0.4588 | 0.052 | 0.0677 | 0.0192 | 0.0219 | 0.0103 | 0.0109 | 0.0064 | 0.0064 |
| | | 25 | 0.4064 | 0.4447 | 0.0336 | 0.0399 | 0.0125 | 0.0125 | 0.0067 | 0.0063 | 0.0041 | 0.0037 |
| Arabidopsis | 0 | 17 | - | 0.9166 | - | 0.0598 | - | 0.0114 | - | 0.0043 | - | 0.0023 |
| | | 25 | - | 0.9673 | - | 0.0216 | - | 0.0046 | - | 0.0021 | - | 0.0012 |
| | 0.0001 | 17 | 0.0033 | 0.9133 | 0.0003 | 0.0594 | 0.0001 | 0.0114 | 0 | 0.0043 | 0 | 0.0023 |
| | | 25 | 0.005 | 0.9624 | 0.0001 | 0.0215 | 0 | 0.0046 | 0 | 0.0021 | 0 | 0.0012 |
| | 0.001 | 17 | 0.0329 | 0.8841 | 0.003 | 0.0566 | 0.0008 | 0.0106 | 0.0004 | 0.0039 | 0.0002 | 0.0021 |
| | | 25 | 0.049 | 0.919 | 0.0011 | 0.02 | 0.0004 | 0.0042 | 0.0002 | 0.0019 | 0.0001 | 0.0011 |
| | 0.01 | 17 | 0.2649 | 0.6575 | 0.0205 | 0.0363 | 0.0043 | 0.0061 | 0.0017 | 0.0021 | 0.0009 | 0.001 |
| | | 25 | 0.3653 | 0.6085 | 0.0066 | 0.0109 | 0.0017 | 0.0021 | 0.0008 | 0.0009 | 0.0005 | 0.0005 |
| Ecoli | 0 | 17 | - | 0.9906 | - | 0.005 | - | 0.0017 | - | 0.0005 | - | 0.0004 |
| | | 25 | - | 0.993 | - | 0.003 | - | 0.0015 | - | 0.0004 | - | 0.0004 |
| | 0.0001 | 17 | 0.0035 | 0.9871 | 0 | 0.005 | 0 | 0.0016 | 0 | 0.0005 | 0 | 0.0004 |
| | | 25 | 0.0051 | 0.9878 | 0 | 0.003 | 0 | 0.0015 | 0 | 0.0004 | 0 | 0.0004 |
| | 0.001 | 17 | 0.0333 | 0.9574 | 0.0002 | 0.0048 | 0.0001 | 0.0015 | 0 | 0.0004 | 0 | 0.0004 |
| | | 25 | 0.0483 | 0.9448 | 0.0001 | 0.0028 | 0.0001 | 0.0014 | 0.0001 | 0.0003 | 0 | 0.0003 |
| | 0.01 | 17 | 0.2717 | 0.72 | 0.0014 | 0.0031 | 0.0005 | 0.0009 | 0.0002 | 0.0002 | 0.0001 | 0.0002 |
| | | 25 | 0.3638 | 0.6305 | 0.001 | 0.0016 | 0.0005 | 0.0007 | 0.0001 | 0.0002 | 0.0002 | 0.0002 |

This table contains the reference $a_i$ values for each genome with various heterozygous rates. Note that only the $a_i$ (i<=5) values are shown in this Table. Here a[i] is equivalent form with $a_i$, and this form is also used in the other Tables of this paper.

**Table S2. Estimation of heterozygous rate using simulated sequencing data.**

| Species | error | k- | Het-rate 0 | Het-rate 0.01% | Het-rate 0.1% | Het-rate 1% |
|---|---|---|---|---|---|---|

| | rate | mer size | a[1/2] | estimated het-rate | a[1/2] | estimated het-rate | a[1/2] | estimated het-rate | a[1/2] | estimated het-rate |
|---|---|---|---|---|---|---|---|---|---|---|
| E.coli_K-12 | 0 | 17 | 0.20% | 0.01% | 0.40% | 0.01% | 3.53% | 0.11% | 27.26% | 0.93% |
| | | 25 | 0.15% | 0.00% | 0.62% | 0.01% | 5.14% | 0.11% | 36.49% | 0.89% |
| | 0.005 | 17 | 0.30% | 0.01% | 0.44% | 0.01% | 3.82% | 0.11% | 27.39% | 0.93% |
| | | 25 | 0.43% | 0.01% | 0.59% | 0.01% | 5.44% | 0.11% | 36.57% | 0.90% |
| | 0.01 | 17 | 0.38% | 0.01% | 0.55% | 0.02% | 3.42% | 0.10% | 27.12% | 0.92% |
| | | 25 | 0.17% | 0.00% | 0.77% | 0.02% | 4.91% | 0.10% | 36.36% | 0.89% |
| Arabidopsis | 0 | 17 | 0.14% | 0.00% | 0.49% | 0.01% | 3.53% | 0.11% | 26.62% | 0.90% |
| | | 25 | 0.13% | 0.00% | 0.63% | 0.01% | 5.07% | 0.10% | 36.48% | 0.89% |
| | 0.005 | 17 | 0.32% | 0.01% | 0.65% | 0.02% | 3.62% | 0.11% | 26.42% | 0.90% |
| | | 25 | 0.29% | 0.01% | 0.72% | 0.02% | 5.04% | 0.10% | 36.53% | 0.89% |
| | 0.01 | 17 | 3.35% | 0.10% | 0.84% | 0.03% | 3.76% | 0.11% | 26.31% | 0.89% |
| | | 25 | 4.37% | 0.09% | 0.97% | 0.02% | 5.17% | 0.11% | 36.43% | 0.89% |
| Human | 0 | 17 | 0.07% | 0.00% | 0.38% | 0.01% | 3.18% | 0.10% | 20.92% | 0.69% |
| | | 25 | 0.12% | 0.00% | 0.63% | 0.01% | 5.60% | 0.12% | 36.20% | 0.88% |
| | 0.005 | 17 | 0.46% | 0.01% | 0.71% | 0.02% | 2.87% | 0.09% | 20.08% | 0.66% |
| | | 25 | 0.46% | 0.01% | 0.91% | 0.02% | 5.28% | 0.11% | 36.60% | 0.90% |
| | 0.01 | 17 | 2.98% | 0.09% | 1.87% | 0.06% | 3.90% | 0.12% | 20.19% | 0.66% |
| | | 25 | 2.32% | 0.05% | 1.50% | 0.03% | 5.74% | 0.12% | 36.33% | 0.89% |
| Maize | 0 | 17 | 0.08% | 0.00% | 0.81% | 0.02% | 4.90% | 0.15% | 31.30% | 1.09% |
| | | 25 | 0.08% | 0.00% | 1.04% | 0.02% | 6.83% | 0.14% | 41.71% | 1.05% |
| | 0.005 | 17 | 3.57% | 0.11% | 3.81% | 0.11% | 7.51% | 0.23% | 30.61% | 1.06% |
| | | 25 | 3.64% | 0.07% | 4.11% | 0.08% | 9.60% | 0.20% | 41.53% | 1.05% |
| | 0.01 | 17 | 6.56% | 0.20% | 6.88% | 0.21% | 9.71% | 0.30% | 30.50% | 1.06% |
| | | 25 | 6.64% | 0.14% | 7.12% | 0.15% | 11.47% | 0.24% | 41.45% | 1.05% |

This table contains the estimation of heterozygous rate using simulated sequencing data. Note that The a[1/2] values were estimated using the heterozygous model, and the estimated heterozygous rate is calculated using a[1/2] values by formula (11) in the main text.

**Table S3. Detailed results of G and a1 estimation with different methods in real data.**

| Species | Error preproccess | rough estimate | | | standard model | | | | continuous model | | | |
|---|---|---|---|---|---|---|---|---|---|---|---|---|
| | | k-mer num | kmer depth | genome size | k-mer num | kmer depth | genome size | a[1] | k-mer num | kmer depth | genome size | a[1] |
| Ant | all | 8,957,619,216 | 26 | 344,523,816 | 8,412,768,579 | 26.99 | 311,699,465 | 82.35% | 8,424,523,576 | 28.08 | 299,968,437 | 83.32% |
| | corrected | 8,420,417,505 | 27 | 311,867,315 | 8,418,025,783 | 27.77 | 303,181,843 | 83.60% | 8,418,520,262 | 27.19 | 309,661,529 | 80.58% |
| | filtered | 6,176,724,320 | 18 | 343,151,351 | 6,161,110,494 | 18.93 | 325,397,589 | 77.29% | 6,163,847,505 | 19.77 | 311,771,512 | 78.40% |
| E.coli | all | 177,595,732 | 32 | 5,549,866 | 170,393,541 | 32.33 | 5,270,968 | 94.96% | 170,519,119 | 32.15 | 5,304,537 | 93.95% |
| | corrected | 173,615,476 | 31 | 5,600,499 | 173,572,790 | 31.98 | 5,428,390 | 93.86% | 173,577,319 | 33.35 | 5,204,046 | 93.59% |
| | filtered | 122,818,870 | 21 | 5,848,517 | 122,456,749 | 21.86 | 5,601,505 | 89.38% | 122,471,962 | 22.68 | 5,400,141 | 88.93% |
| Human | all | 84,840,631,322 | 28 | 3,030,022,547 | 83,726,746,695 | 28.77 | 2,909,815,725 | 66.37% | 83,797,402,710 | 28.08 | 2,983,982,946 | 65.12% |

| | | | | | | | | | | | | | |
|---|---|---|---|---|---|---|---|---|---|---|---|---|---|
| | corrected | 82,507,739,275 | 28 | 2,946,704,974 | 82,498,370,101 | 28.74 | 2,870,287,246 | 66.90% | 82,500,224,647 | 28 | 2,946,741,792 | 65.31% |
| | filtered | 67,127,996,412 | 22 | 3,051,272,564 | 67,033,584,827 | 22.97 | 2,918,462,646 | 66.00% | 67,041,928,286 | 23.52 | 2,850,470,598 | 67.40% |
| Panda | all | 71,779,930,870 | 28 | 2,563,568,959 | 67,652,851,884 | 28.82 | 2,347,329,461 | 72.21% | 67,777,715,418 | 28.28 | 2,396,293,201 | 70.63% |
| | corrected | 67,461,715,871 | 28 | 2,409,346,995 | 67,357,673,680 | 28.88 | 2,332,652,503 | 72.14% | 67,375,272,401 | 28.42 | 2,370,340,602 | 70.35% |
| | filtered | 41,137,783,154 | 16 | 2,571,111,447 | 40,800,034,010 | 16.89 | 2,415,518,152 | 71.30% | 40,813,266,099 | 17.54 | 2,326,443,641 | 72.16% |
| Potato | all | 22,403,643,321 | 26 | 861,678,589 | 21,143,648,108 | 26.71 | 791,473,035 | 75.61% | 21,185,080,846 | 26.18 | 809,264,229 | 73.55% |
| | corrected | 21,424,465,114 | 27 | 793,498,707 | 21,405,809,036 | 27.62 | 775,067,312 | 76.05% | 21,407,775,752 | 27.22 | 786,567,649 | 73.85% |
| | filtered | 14,136,873,771 | 17 | 831,580,810 | 13,994,215,770 | 16.99 | 823,523,416 | 73.65% | 14,000,540,319 | 17.19 | 814,396,835 | 74.59% |

This table contains the detailed results of G and a1 estimation with different methods in real data. Note that there are three data types: "all" means all the sequencing data were used to count k-mer frequency; "corrected" means the data were error corrected before counting k-mer frequency; "filtered" means all the raw sequencing data were used to count k-mer frequency, but the k-mers with low quality were filtered. The detailed descriptions for all the estimation methods can be found in the main text.

**Table S4. The $c_k$ and $a_k$ values for real sequencing data by the continuous model.**

| | Ant | | E.coli | | Human | | Panda | | Potato | |
|---|---|---|---|---|---|---|---|---|---|---|
| $k$ | $a_k$ | $c_k$ | $a_k$ | $c_k$ | $a_k$ | $c_k$ | $a_k$ | $c_k$ | $a_k$ | $c_k$ |
| 1 | 0.00% | 0.00 | 0.00% | 0.00 | 0.00% | 0.00 | 0.00% | 0.00 | 0.00% | 0.00 |
| 2 | 0.00% | 10.60 | 0.00% | 12.19 | 0.00% | 11.35 | 0.00% | 12.85 | 0.00% | 12.78 |
| 3 | 0.14% | 10.78 | 0.07% | 12.21 | 0.07% | 11.39 | 0.03% | 12.90 | 0.08% | 12.94 |
| 4 | 2.79% | 11.22 | 2.59% | 12.57 | 3.37% | 11.52 | 2.24% | 13.00 | 2.13% | 13.22 |
| 5 | 5.62% | 15.45 | 8.00% | 19.61 | 4.93% | 16.24 | 5.09% | 15.69 | 5.96% | 14.90 |
| 6 | 10.47% | 20.31 | 13.14% | 23.78 | 8.73% | 21.29 | 9.17% | 21.97 | 10.14% | 20.36 |
| 7 | 13.88% | 23.21 | 17.65% | 28.22 | 11.40% | 24.56 | 13.49% | 24.82 | 13.92% | 23.21 |
| 8 | 15.27% | 26.52 | 19.53% | 32.15 | 12.01% | 28.27 | 14.98% | 28.54 | 15.21% | 26.30 |
| 9 | 14.20% | 29.76 | 16.48% | 35.24 | 10.80% | 31.59 | 13.22% | 31.53 | 13.05% | 29.21 |
| 10 | 11.39% | 32.66 | 10.50% | 38.53 | 8.62% | 34.74 | 8.51% | 34.12 | 9.21% | 31.74 |
| 11 | 7.05% | 35.43 | 5.67% | 42.73 | 5.80% | 38.18 | 5.07% | 37.36 | 5.47% | 34.50 |
| 12 | 4.23% | 38.45 | 2.71% | 46.35 | 3.85% | 42.12 | 2.85% | 42.32 | 2.70% | 38.47 |
| 13 | 2.46% | 42.12 | 1.15% | 49.42 | 3.04% | 46.30 | 2.38% | 47.34 | 1.89% | 43.46 |
| 14 | 1.34% | 46.50 | 0.47% | 55.35 | 2.54% | 50.33 | 2.30% | 51.09 | 1.66% | 47.48 |
| 15 | 1.01% | 50.95 | 0.32% | 62.96 | 2.42% | 54.04 | 2.38% | 54.28 | 1.59% | 50.60 |
| 16 | 0.89% | 54.81 | 0.25% | 65.76 | 2.29% | 57.48 | 2.37% | 57.35 | 1.58% | 53.42 |
| 17 | 0.85% | 58.06 | 0.21% | 67.87 | 2.13% | 60.78 | 2.15% | 60.46 | 1.50% | 56.26 |
| 18 | 0.84% | 60.95 | 0.17% | 71.02 | 1.96% | 64.09 | 1.86% | 63.67 | 1.31% | 59.23 |
| 19 | 0.79% | 63.74 | 0.14% | 76.95 | 1.68% | 67.52 | 1.45% | 67.07 | 1.11% | 62.37 |
| 20 | 0.70% | 66.59 | 0.12% | 81.41 | 1.47% | 71.07 | 1.11% | 70.80 | 0.92% | 65.66 |
| 21 | 0.60% | 69.63 | 0.10% | 83.70 | 1.22% | 74.75 | 0.92% | 74.95 | 0.72% | 69.14 |
| 22 | 0.47% | 72.92 | 0.08% | 86.13 | 1.03% | 78.50 | 0.77% | 79.17 | 0.61% | 72.81 |
| 23 | 0.38% | 76.46 | 0.05% | 92.22 | 0.91% | 82.26 | 0.71% | 83.04 | 0.54% | 76.56 |
| 24 | 0.32% | 80.23 | 0.05% | 99.46 | 0.79% | 85.97 | 0.64% | 86.47 | 0.48% | 80.17 |
| 25 | 0.27% | 84.10 | 0.04% | 101.84 | 0.73% | 89.58 | 0.58% | 89.66 | 0.46% | 83.52 |
| 26 | 0.24% | 87.84 | 0.04% | 103.32 | 0.64% | 93.12 | 0.52% | 92.85 | 0.43% | 86.65 |
| 27 | 0.22% | 91.29 | 0.03% | 105.35 | 0.57% | 96.63 | 0.45% | 96.24 | 0.39% | 89.69 |
| 28 | 0.21% | 94.48 | 0.02% | 112.09 | 0.52% | 100.17 | 0.39% | 99.92 | 0.36% | 92.76 |
| 29 | 0.19% | 97.53 | 0.02% | 120.53 | 0.46% | 103.78 | 0.34% | 103.84 | 0.33% | 95.97 |
| 30 | 0.18% | 100.60 | 0.02% | 122.71 | 0.42% | 107.42 | 0.29% | 107.78 | 0.29% | 99.39 |
| 31 | 0.16% | 103.80 | 0.02% | 123.76 | 0.38% | 111.04 | 0.27% | 111.54 | 0.26% | 102.99 |
| 32 | 0.15% | 107.19 | 0.01% | 124.69 | 0.33% | 114.60 | 0.24% | 115.06 | 0.24% | 106.64 |
| 33 | 0.13% | 110.74 | 0.01% | 126.26 | 0.31% | 118.15 | 0.22% | 118.45 | 0.22% | 110.17 |
| 34 | 0.12% | 114.36 | 0.01% | 136.88 | 0.27% | 121.79 | 0.20% | 121.89 | 0.21% | 113.52 |
| 35 | 0.11% | 117.96 | 0.01% | 149.27 | 0.25% | 125.61 | 0.18% | 125.54 | 0.20% | 116.73 |
| 36 | 0.10% | 121.53 | 0.01% | 151.18 | 0.23% | 129.58 | 0.16% | 129.49 | 0.19% | 119.95 |
| 37 | 0.10% | 125.10 | 0.01% | 152.17 | 0.21% | 133.49 | 0.15% | 133.55 | 0.18% | 123.37 |
| 38 | 0.09% | 128.71 | 0.01% | 153.03 | 0.21% | 137.16 | 0.14% | 137.41 | 0.17% | 127.09 |
| 39 | 0.09% | 132.38 | 0.01% | 154.14 | 0.19% | 140.57 | 0.13% | 140.94 | 0.16% | 131.07 |
| 40 | 0.09% | 136.08 | 0.01% | 156.62 | 0.18% | 143.99 | 0.12% | 144.34 | 0.16% | 135.07 |
| 41 | 0.08% | 139.86 | 0.01% | 169.81 | 0.17% | 147.96 | 0.11% | 148.16 | 0.16% | 138.88 |
| 42 | 0.09% | 143.92 | 0.01% | 175.48 | 0.16% | 153.47 | 0.11% | 153.39 | 0.16% | 142.64 |
| 43 | 0.09% | 148.77 | 0.01% | 176.35 | 0.17% | 160.58 | 0.11% | 160.47 | 0.16% | 147.03 |
| 44 | 0.09% | 155.56 | 0.01% | 200.84 | 0.19% | 165.80 | 0.13% | 166.00 | 0.17% | 153.78 |
| 45 | 0.12% | 164.11 | 0.02% | 200.84 | 0.22% | 168.92 | 0.14% | 169.25 | 0.23% | 163.43 |
| 46 | 0.17% | 171.73 | 0.03% | 200.84 | 0.16% | 178.70 | 0.11% | 177.39 | 0.33% | 170.11 |
| 47 | 0.28% | 197.35 | 0.01% | 246.35 | 0.50% | 198.30 | 0.33% | 198.36 | 0.53% | 196.40 |
| 48 | 0.96% | 250.58 | 0.15% | 246.35 | 1.45% | 250.07 | 0.90% | 249.91 | 2.24% | 251.20 |

This file contains the $c_k$ and $a_k$ values for real sequencing data by the continuous model. Note that 48 ranks (k) were designed originally to represent i up to 6, with each i has 8 ranks.

**Table S5. Comparison of estimation accuracy between GSP and GCE**

| #Species | kmer_size | error | snp | gsp_estimate_genome_size | gsp_a[1] | gce_estimate_genome_size | gce_estimate_a[1] | gsp_ΔG/G | gce_ΔG/G | gsp_Δa[1]/a[1] | gce |
|---|---|---|---|---|---|---|---|---|---|---|---|
| Arabidopsis | 17 | 0.E+00 | 0.E+00 | 113993564 | 0.92 | 119182048 | 0.92 | 4.21% | 0.15% | 0.70% | |
| Arabidopsis | 17 | 5.E-03 | 0.E+00 | 114030797 | 0.37 | 119227035 | 0.92 | 4.17% | 0.19% | 59.31% | |
| Arabidopsis | 17 | 1.E-02 | 0.E+00 | 115279632 | 0.27 | 119479778 | 0.90 | 3.12% | 0.41% | 70.98% | |
| Arabidopsis | 25 | 0.E+00 | 0.E+00 | 116529243 | 0.97 | 119133663 | 0.97 | 2.07% | 0.11% | 0.18% | |
| Arabidopsis | 25 | 5.E-03 | 0.E+00 | 116630407 | 0.33 | 119112889 | 0.97 | 1.99% | 0.10% | 65.47% | |
| Arabidopsis | 25 | 1.E-02 | 0.E+00 | 117901514 | 0.23 | 119781621 | 0.95 | 0.92% | 0.66% | 76.22% | |
| Ecoli | 17 | 0.E+00 | 0.E+00 | 4632494 | 0.99 | 4654689 | 0.99 | 0.15% | 0.32% | 0.06% | |
| Ecoli | 17 | 5.E-03 | 0.E+00 | 4677692 | 0.40 | 4652932 | 0.99 | 0.82% | 0.29% | 59.42% | |
| Ecoli | 17 | 1.E-02 | 0.E+00 | 4635012 | 0.27 | 4633878 | 0.99 | 0.10% | 0.12% | 72.84% | |
| Ecoli | 25 | 0.E+00 | 0.E+00 | 4635069 | 0.99 | 4667422 | 0.99 | 0.10% | 0.60% | 0.00% | |
| Ecoli | 25 | 5.E-03 | 0.E+00 | 4634535 | 0.36 | 4626684 | 0.99 | 0.11% | 0.28% | 64.25% | |
| Ecoli | 25 | 1.E-02 | 0.E+00 | 4637042 | 0.23 | 4690747 | 0.99 | 0.06% | 1.10% | 76.74% | |

| | | | | | | | | | | |
|---|---|---|---|---|---|---|---|---|---|---|
| Human | 17 | 0.E+00 | 0.E+00 | 1747267956 | 0.86 | 2836465136 | 0.67 | 38.31% | 0.14% | 26.59% |
| Human | 17 | 5.E-03 | 0.E+00 | 1749246652 | 0.43 | 2831972523 | 0.67 | 38.24% | 0.01% | 35.97% |
| Human | 17 | 1.E-02 | 0.E+00 | 1762593558 | 0.35 | 2848494229 | 0.67 | 37.77% | 0.57% | 48.98% |
| Human | 25 | 0.E+00 | 0.E+00 | 2492726920 | 0.98 | 2834895501 | 0.97 | 11.99% | 0.09% | 0.89% |
| Human | 25 | 5.E-03 | 0.E+00 | 2501613901 | 0.34 | 2857474431 | 0.97 | 11.68% | 0.89% | 65.06% |
| Human | 25 | 1.E-02 | 0.E+00 | 2516523947 | 0.23 | 2896908560 | 0.96 | 11.15% | 2.28% | 76.43% |
| Maize | 17 | 0.E+00 | 0.E+00 | 518557098 | 0.95 | 2037685440 | 0.73 | 74.50% | 0.21% | 29.15% |
| Maize | 17 | 5.E-03 | 0.E+00 | 535682479 | 0.31 | 2058430932 | 0.72 | 73.66% | 1.23% | 57.77% |
| Maize | 17 | 1.E-02 | 0.E+00 | 571667131 | 0.22 | 2116150329 | 0.71 | 71.89% | 4.07% | 70.21% |
| Maize | 25 | 0.E+00 | 0.E+00 | 789605809 | 0.94 | 2035563354 | 0.81 | 61.17% | 0.10% | 15.56% |
| Maize | 25 | 5.E-03 | 0.E+00 | 835513043 | 0.24 | 2084690632 | 0.81 | 58.91% | 2.52% | 70.04% |
| Maize | 25 | 1.E-02 | 0.E+00 | 871713364 | 0.16 | 2161099906 | 0.80 | 57.13% | 6.28% | 80.84% |
| Arabidopsis | 17 | 0.E+00 | 1.E-04 | 114223117 | 0.92 | 119433737 | 0.92 | 4.01% | 0.37% | 0.70% |
| Arabidopsis | 17 | 5.E-03 | 1.E-04 | 114226522 | 0.37 | 119141613 | 0.92 | 4.01% | 0.12% | 59.31% |
| Arabidopsis | 17 | 1.E-02 | 1.E-04 | 114378568 | 0.25 | 119243299 | 0.92 | 3.88% | 0.21% | 72.62% |
| Arabidopsis | 25 | 0.E+00 | 1.E-04 | 228266169 | 0.02 | 119230332 | 0.97 | 91.82% | 0.20% | 97.83% |
| Arabidopsis | 25 | 5.E-03 | 1.E-04 | 116956240 | 0.33 | 119109919 | 0.97 | 1.72% | 0.09% | 65.47% |
| Arabidopsis | 25 | 1.E-02 | 1.E-04 | 117161156 | 0.22 | 119118473 | 0.97 | 1.54% | 0.10% | 77.67% |
| Ecoli | 17 | 0.E+00 | 1.E-04 | 4640150 | 0.99 | 4682725 | 0.99 | 0.01% | 0.93% | 0.06% |
| Ecoli | 17 | 5.E-03 | 1.E-04 | 4641574 | 0.41 | 4665487 | 0.99 | 0.04% | 0.56% | 58.81% |
| Ecoli | 17 | 1.E-02 | 1.E-04 | 4640646 | 0.27 | 4641926 | 0.99 | 0.02% | 0.05% | 72.84% |
| Ecoli | 25 | 0.E+00 | 1.E-04 | 4646552 | 0.99 | 4648897 | 0.99 | 0.15% | 0.20% | 0.00% |
| Ecoli | 25 | 5.E-03 | 1.E-04 | 4647218 | 0.36 | 4659834 | 0.99 | 0.16% | 0.43% | 64.05% |
| Ecoli | 25 | 1.E-02 | 1.E-04 | 4645871 | 0.23 | 4620763 | 0.99 | 0.13% | 0.41% | 76.74% |
| Human | 17 | 0.E+00 | 1.E-04 | 1750534529 | 0.86 | 2838372887 | 0.67 | 38.20% | 0.21% | 26.59% |
| Human | 17 | 5.E-03 | 1.E-04 | 1751831553 | 0.43 | 2830714668 | 0.68 | 38.15% | 0.06% | 35.97% |
| Human | 17 | 1.E-02 | 1.E-04 | 1760548923 | 0.34 | 2841008485 | 0.67 | 37.84% | 0.31% | 50.01% |
| Human | 25 | 0.E+00 | 1.E-04 | 2499463943 | 0.98 | 2835970370 | 0.97 | 11.75% | 0.13% | 0.89% |
| Human | 25 | 5.E-03 | 1.E-04 | 2507460686 | 0.34 | 2856020286 | 0.97 | 11.47% | 0.84% | 65.06% |
| Human | 25 | 1.E-02 | 1.E-04 | 2520527244 | 0.22 | 2885870063 | 0.97 | 11.01% | 1.89% | 77.36% |
| Maize | 17 | 0.E+00 | 1.E-04 | 520388042 | 0.95 | 2041728303 | 0.73 | 74.41% | 0.41% | 29.15% |
| Maize | 17 | 5.E-03 | 1.E-04 | 536702116 | 0.31 | 2065464402 | 0.72 | 73.61% | 1.57% | 57.77% |
| Maize | 17 | 1.E-02 | 1.E-04 | 572587757 | 0.22 | 2116256118 | 0.71 | 71.84% | 4.07% | 70.21% |
| Maize | 25 | 0.E+00 | 1.E-04 | 793501673 | 0.94 | 2040454389 | 0.81 | 60.98% | 0.34% | 15.56% |
| Maize | 25 | 5.E-03 | 1.E-04 | 838042754 | 0.24 | 2086280421 | 0.81 | 58.79% | 2.60% | 70.04% |
| Maize | 25 | 1.E-02 | 1.E-04 | 873744098 | 0.16 | 2161275222 | 0.80 | 57.03% | 6.28% | 80.72% |
| Arabidopsis | 17 | 0.E+00 | 1.E-03 | 116178157 | 0.92 | 119469709 | 0.87 | 2.37% | 0.40% | 0.70% |
| Arabidopsis | 17 | 5.E-03 | 1.E-03 | 116126936 | 0.38 | 119205642 | 0.87 | 2.41% | 0.17% | 58.98% |
| Arabidopsis | 17 | 1.E-02 | 1.E-03 | 116483834 | 0.25 | 119655679 | 0.87 | 2.11% | 0.55% | 72.29% |
| Arabidopsis | 25 | 0.E+00 | 1.E-03 | 119778936 | 0.97 | 119284430 | 0.91 | 0.66% | 0.24% | 0.07% |
| Arabidopsis | 25 | 5.E-03 | 1.E-03 | 119694666 | 0.34 | 118930497 | 0.91 | 0.59% | 0.06% | 65.06% |
| Arabidopsis | 25 | 1.E-02 | 1.E-03 | 119868482 | 0.22 | 119258955 | 0.91 | 0.73% | 0.22% | 77.36% |
| Ecoli | 17 | 0.E+00 | 1.E-03 | 4714392 | 0.99 | 4664276 | 0.95 | 1.61% | 0.53% | 0.06% |
| Ecoli | 17 | 5.E-03 | 1.E-03 | 4712400 | 0.41 | 4631437 | 0.94 | 1.57% | 0.18% | 58.61% |
| Ecoli | 17 | 1.E-02 | 1.E-03 | 4711747 | 0.27 | 4644340 | 0.95 | 1.55% | 0.10% | 72.54% |
| Ecoli | 25 | 0.E+00 | 1.E-03 | 4754346 | 0.99 | 4637500 | 0.93 | 2.47% | 0.05% | 0.10% |
| Ecoli | 25 | 5.E-03 | 1.E-03 | 4755820 | 0.36 | 4612603 | 0.93 | 2.50% | 0.58% | 63.65% |
| Ecoli | 25 | 1.E-02 | 1.E-03 | 4743675 | 0.24 | 4642938 | 0.94 | 2.24% | 0.07% | 76.33% |
| Human | 17 | 0.E+00 | 1.E-03 | 1781877085 | 0.85 | 2852338189 | 0.62 | 37.09% | 0.71% | 26.00% |
| Human | 17 | 5.E-03 | 1.E-03 | 1775011701 | 0.44 | 2840889922 | 0.63 | 37.33% | 0.30% | 35.67% |
| Human | 17 | 1.E-02 | 1.E-03 | 1779351166 | 0.34 | 2859731024 | 0.62 | 37.18% | 0.97% | 49.87% |
| Human | 25 | 0.E+00 | 1.E-03 | 2560242878 | 0.98 | 2850753099 | 0.89 | 9.61% | 0.65% | 0.79% |
| Human | 25 | 5.E-03 | 1.E-03 | 2568748123 | 0.34 | 2862009236 | 0.91 | 9.31% | 1.05% | 64.54% |
| Human | 25 | 1.E-02 | 1.E-03 | 2586399277 | 0.22 | 2888544012 | 0.90 | 8.68% | 1.98% | 76.95% |
| Maize | 17 | 0.E+00 | 1.E-03 | 536585544 | 0.95 | 2039037547 | 0.68 | 73.61% | 0.27% | 29.15% |
| Maize | 17 | 5.E-03 | 1.E-03 | 551339418 | 0.32 | 2079155376 | 0.63 | 72.89% | 2.25% | 56.68% |
| Maize | 17 | 1.E-02 | 1.E-03 | 580792069 | 0.22 | 2132256530 | 0.60 | 71.44% | 4.86% | 70.07% |
| Maize | 25 | 0.E+00 | 1.E-03 | 827749160 | 0.94 | 2037667192 | 0.74 | 59.29% | 0.21% | 15.56% |
| Maize | 25 | 5.E-03 | 1.E-03 | 860151082 | 0.25 | 2098695048 | 0.70 | 57.70% | 3.21% | 69.42% |
| Maize | 25 | 1.E-02 | 1.E-03 | 904636916 | 0.16 | 2174457885 | 0.68 | 55.51% | 6.93% | 80.60% |
| Arabidopsis | 17 | 0.E+00 | 1.E-02 | 137172546 | 0.90 | 119502846 | 0.65 | 15.27% | 0.42% | 2.25% |
| Arabidopsis | 17 | 5.E-03 | 1.E-02 | 137057183 | 0.40 | 119438723 | 0.65 | 15.18% | 0.37% | 56.91% |
| Arabidopsis | 17 | 1.E-02 | 1.E-02 | 138144123 | 0.27 | 120170377 | 0.65 | 16.09% | 0.99% | 70.43% |
| Arabidopsis | 25 | 0.E+00 | 1.E-02 | 218874698 | 0.45 | 119464058 | 0.60 | 83.93% | 0.39% | 53.69% |
| Arabidopsis | 25 | 5.E-03 | 1.E-02 | 217156887 | 0.18 | 118803053 | 0.60 | 82.49% | 0.16% | 81.29% |
| Arabidopsis | 25 | 1.E-02 | 1.E-02 | 216070832 | 0.12 | 119285241 | 0.60 | 81.58% | 0.24% | 87.28% |
| Ecoli | 17 | 0.E+00 | 1.E-02 | 5615290 | 0.94 | 4653027 | 0.71 | 21.03% | 0.29% | 4.81% |
| Ecoli | 17 | 5.E-03 | 1.E-02 | 5603955 | 0.42 | 4636462 | 0.71 | 20.78% | 0.07% | 57.40% |
| Ecoli | 17 | 1.E-02 | 1.E-02 | 5699551 | 0.28 | 4650758 | 0.72 | 22.84% | 0.24% | 71.63% |
| Ecoli | 25 | 0.E+00 | 1.E-02 | 9133062 | 0.38 | 4643339 | 0.63 | 96.85% | 0.08% | 61.73% |
| Ecoli | 25 | 5.E-03 | 1.E-02 | 9114588 | 0.16 | 4609669 | 0.63 | 96.45% | 0.65% | 84.39% |
| Ecoli | 25 | 1.E-02 | 1.E-02 | 9128281 | 0.10 | 4597866 | 0.63 | 96.74% | 0.90% | 89.53% |
| Human | 17 | 0.E+00 | 1.E-02 | 2054684008 | 0.83 | 2865254085 | 0.45 | 27.46% | 1.16% | 22.89% |
| Human | 17 | 5.E-03 | 1.E-02 | 2047915840 | 0.46 | 2852169971 | 0.47 | 27.70% | 0.70% | 32.42% |
| Human | 17 | 1.E-02 | 1.E-02 | 2061230383 | 0.37 | 2908235720 | 0.46 | 27.23% | 2.68% | 45.58% |
| Human | 25 | 0.E+00 | 1.E-02 | 3129041400 | 0.96 | 2874261953 | 0.59 | 10.47% | 1.48% | 0.56% |
| Human | 25 | 5.E-03 | 1.E-02 | 3123422298 | 0.38 | 2864463640 | 0.60 | 10.28% | 1.13% | 60.51% |
| Human | 25 | 1.E-02 | 1.E-02 | 3183400262 | 0.26 | 2930637221 | 0.59 | 12.39% | 3.47% | 72.71% |
| Maize | 17 | 0.E+00 | 1.E-02 | 695819120 | 0.94 | 2046699465 | 0.45 | 65.78% | 0.65% | 28.60% |
| Maize | 17 | 5.E-03 | 1.E-02 | 719596150 | 0.37 | 2096864740 | 0.45 | 64.61% | 3.12% | 49.98% |
| Maize | 17 | 1.E-02 | 1.E-02 | 752269146 | 0.26 | 2162585546 | 0.44 | 63.01% | 6.35% | 64.47% |
| Maize | 25 | 0.E+00 | 1.E-02 | 1160175379 | 0.94 | 2047298540 | 0.44 | 42.95% | 0.68% | 15.19% |
| Maize | 25 | 5.E-03 | 1.E-02 | 1202324526 | 0.30 | 2111723426 | 0.44 | 40.87% | 3.85% | 62.67% |
| Maize | 25 | 1.E-02 | 1.E-02 | 1202098973 | 0.19 | 2198481293 | 0.43 | 40.88% | 8.11% | 76.42% |

This table contains the detailed results for GSP and GCE estimation results, including the genome size, a1 values, and all the middle results.

# Supplementary methods

## 1 k-mer frequency counting

Counting k-mer frequency from either the reference genome sequence or sequenced reads data is one of the most fundamental analyses in bioinformatics. To estimate genomic characters, the k-mer size ($K$) should be determined under the logic that the space of k-mers ($4^K$) should be several times larger than the genome size ($G$), so that few k-mers derived from different genomic positions will merge together by chance, i.e. most k-mers in the genome will appear uniquely. In practice, we often require the k-mer space to be at least 5 times larger than the genome size ($4^K > 5*G$), and the larger the better. When the k-mer size for a special application is determined, the next question is to find an appropriate tool to count k-mer frequency. From our experience, the recently published jellyfish algorithm [1] performs better than the others, which has relatively faster running speed and smaller memory requirement, as well as broader accessible range of k-mer size (up to 31 bp). Note that the computer memory consumption for k-mer frequency counting is often very large, especially for data from large genome and which has many sequencing errors, so you'd better pre-estimate the memory usage and find a computer with enough memory to run the task. When the estimated memory usage exceeds that of the largest-memory computer you can access, then the only way is to choose or design some memory saving algorithms such as bloom filter [2]. We do not discuss more details on k-mer counting algorithms here, because it is not the focus of this paper.

We also developed a counting k-mer program named as kmerfreq, which contains two methods: "kmer_freq_array" uses the array index to represent k-mers, and requires $4^K$ computer memory for any data; "kmer_freq_hash" uses hash key to store k-mers, although it requires much more memory to store one k-mer, however the memory requirement is in line with the number of k-mer species in the given data. For smaller k-mer size ($K$<=17), it is better to use "kmer_freq_array", while for larger k-mer size ($K$<=27), "kmer_freq_hash" will be the preferred.

# 2 Estimation of the expected coverage depth of k-mer ($c$)

## 2.1 Estimating $c$ based on definition

For genomes without repeat, it is defined as that:

$$c = \sum xp(x) = \frac{n_{Kindividuals}}{n_{Kspecies}} = \frac{n_{Kindividuals}}{G}, \quad n_{Kspecies} = G$$

For genomes with repeat, it can be calculated by:

$$c = \frac{\sum xp(x)}{\sum i * a_i} = \frac{n_{Kindividuals} / n_{Kspecies}}{G / n_{Kspecies}} = \frac{n_{Kindividuals}}{G}$$

## 2.2 Estimating $c$ from Poisson distribution

For genomes without repeat, $P_{Kspecies}(x)$ follows the Poisson distribution. Then:

$$P_{Kspecies}(x) = \frac{c^x}{x!} e^{-c}$$

$$P_{Kindividuals}(x) = xP_{Kspecies}(x)/c = \frac{xc^x}{x!c}e^{-c} = \frac{c^{(x-1)}}{(x-1)!}e^{-c}$$

By deduction, we can get:

$$\frac{P_{Kspecies}(x)}{P_{Kspecies}(x+1)} = \frac{\frac{c^x}{x!}e^{-c}}{\frac{c^{x+1}}{(x+1)!}e^{-c}} = \frac{x+1}{c}$$

$$\frac{P_{Kindividuals}(x)}{P_{Kindividuals}(x+1)} = \frac{\frac{c^{x-1}}{(x-1)!}e^{-c}}{\frac{c^x}{x!}e^{-c}} = \frac{x}{c}$$

Thus, we can estimate $c$ based on the following formula:

$$c = \frac{P_{Kspecies}(x+1)}{P_{Kspecies}(x)} \times (x+1)$$

$$c = \frac{P_{Kindividuals}(x+1)}{P_{Kindividuals}(x)} \times x$$

# 3 Exploring the repetitive genomes

The real genomes often contain huge amount of repeat sequences, including transposable elements [3], tandem repeats [4], and segmental duplications [5], which bring great challenge in the assembly processes [6]. The $a_i$ and $b_i$ values calculated by genomic k-mer frequency and the k-mer species and individuals curves plotted by coverage depth, are closely related with repeats in the genome. We can make a rough inference that the genome contains repeats if we see multiple peaks on the k-mer species and individuals curves. To be more quantitative and intuitionistic to reflect genomic repeat character, it is better to use the $a_i$ and $b_i$ values at the same time.

Calculating genome size for repetitive genomes based on the compound Poisson model:

$$G = \sum_{i=1}^{m} i \times n_{i,genomic,Kspecies} = \sum_{i=1}^{m} i \times \frac{n_{i,Kindividuals}}{c_i} = \frac{1}{c} \times \sum_{i=1}^{m} n_{i,Kindividuals} = \frac{n_{Kindividuals}}{c} = \frac{n_{kmer}}{c}$$

Note: For the number of k-mer species and individuals, if not specified as “genomic”, it means in the sequencing data.

# 4 The standard model

For genome with various copies of repeat families, we can divide the genome sequence into $m$ repeat families, and let $i \in [1,m]$ be the copy number of the *ith* repeat families (also the genomic frequency of contained k-mers). We define $a_i$ as the ratio (probability) of k-mer species with genomic frequency $i$ divided by total k-mer species and $b_i$ as the ratio (probability) of k-mer individuals with genomic frequency $i$ divided by total k-mer individuals. Then the whole genome will be sequenced randomly, and the k-mer frequency will be counted from the sequenced reads. There is a coverage depth character for each sequenced k-mer,which is ranging from 1 to $w$. For each coverage depth, we define $v_j$ as the ratio of *jth* depth k-mer species divided by total k-mer species and $u_j$ as the ratio of *jth* depth k-mer individuals divided by total k-mer individuals. It is supposed that the sequencing on each repeat part is independent, so for the k-mers with same genomic frequency $i$, the depth distribution is also following a Poisson distribution. Then we can use the following compound model to describe the total k-mer depth distribution:

$$P_{Kspecies}(x) = \sum_{i=1}^{m} a_i \times P_{Kspecies,i}(x)$$

$$P_{Kindividuals}(x) = \sum_{i=1}^{m} b_i \times P_{Kindividuals,i}(x)$$

Because $i$ is integer and stands for the copy number of the k-mer in the genome, we can get the following formula:

$c_i = i \times c$, $c$ is the k-mer depth of the unique region.

$$G = \sum_{i=1}^{m} i \times \frac{n_{i,Kindividuals}}{c_i} = \sum_{i=1}^{m} i \times \frac{n_{i,Kindividuals}}{i \times c} = \frac{1}{c} \times \sum_{i=1}^{m} n_{i,Kindividuals} = \frac{n_{Kindividuals}}{c}$$

The relation between $a_i$ and $b_i$ is:

$$b_i = \frac{n_{Kspecies} \times a_i \times c_i}{n_{k-mer}}$$

Because there is relation between $a_i$ and $b_i$, we just estimate $a_i$ in the bellow.

To estimate $a_i$, we have the prior probabilities:

$$P(i) = a_i, \quad P(j \mid i) = \frac{(ic)^j e^{-ic}}{j!}$$

$$P(i \mid j) = \frac{P(i)P(j \mid i)}{\sum_{i=1}^{m} P(i)P(j \mid i)} = \frac{a_i P(j \mid i)}{\sum_{i=1}^{m} a_i P(j \mid i)}$$

The posterior probabilities can be calculated by:

$$a_i = \sum_{j=0}^{w} P(i \mid j) \times v_j$$

Then we get the iteration formula:

$$a_{i,t+1} = \sum_{j=0}^{w} \frac{a_{i,t} P(j \mid i)}{\sum_{i=1}^{m} a_{i,t} P(j \mid i)} \times v_j$$

$$c_{t+1} = \frac{P_{Kspecies,1,t+1}(x+1)}{P_{Kspecies,1,t+1}(x)} \times (x+1)$$

$$c_{i,t+1} = i \times c_{t+1}$$

In each iteration cycle, we update $a_i$ and remove the repeat parts from the raw depth distribution curve, then only the unique curve is left and used to estimate the unique depth *c* and update $c_i$.

When *j*=0, the $v_0$ can not be counted directly from reads, which stands for the uncovered region in the genome. Although the sequencing depth is quite high for short read sequencing, there are still some regions not covered by reads. Here we still use the Lander-Waterman model in 1988 to estimate the uncovered gap region. For each *ith* repeat region, the expected k-mer depth is $c_i$, then the gap probability is $e^{-c_i}$, so we can estimate the $v_0$ by:

$$v_0 = \sum_{i=1}^{m} a_i e^{-ic}$$

# 5 Exploring the heterozygous genomes

For heterozygous genomes, the heterozygous sites may cause a new set of peaks on the k-mer coverage depth distribution curves with the peak position at *i***c*/2 named as hybrid peaks. For non-repetitive genome, when the heterozygous rate gets higher, another peak rises at the *c/*2 position. Moreover, when the heterozygous rate passes a threshold, the *c/*2 peak will become the major peak replacing that at depth *c*. It also shows that the two curves have different advantage for different purposes. It is easier to find out the homozygous peak (*c*) in the k-mer individual curve when the heterozygous rate is relatively high, while it is easier to find out the heterozygous peak (*c/*2) in k-mer species curve when the heterozygous rate is relatively low.

We developed the heterozygous discrete model for heterozygous genome, in which *i* is integer times of 1/2. In practice, the height of hybrid peaks may increase as the heterozygous rate gets higher and it might be difficult to distinguish the repeat peaks and the hybrid peaks. Then we need additional biological background of the sequencing sample to make a correct determination.

Heterozygous rate is a quite important character for genome sequencing projects. For non-repetitive genome, when the $4^K$ is much larger than the genome size, one heterozygous site will cause *K* new k-mers and 2**K* k-mers with coverage depth *c*/2, then we can get the formula (11) to roughly estimate the heterozygous rate for diploid genome. For repetitive genome, there is no clear relationship between the $a_i$ and heterozygous rate, so it is difficult to make a formula. However, in practice, we can still use formula (11) to roughly estimate the heterozygous rate for diploid genome with repeats, although it is not as accurate as that of non-repetitive genome.

## 6 Sequencing error model

For reads data with sequencing error, we can divide the erroneous k-mer into two kinds. The first kind of erroneous k-mer has low depth and do not exist in the genome. They have the expected depth $c_{error,out}$. The other kind erroneous k-mers exist in the genome and merge with the correct k-mers increasing their expected depth. Let $c_{error,in}$ be the increment, then we can get the compound model:

$$P_{Kspecies}(x) = \sum_{i=1}^{m} A_i \times P_{Kspecies,i}(x) + \frac{4^k - n_{Kspecies}}{4^k} P_{Kspecies,error}(x)$$

$$A_i = \frac{n_{Kspecies}}{4^k} \times a_i$$

$$P_{Kspecies,i}(x) = \text{Poisson}(c + c_{error,in})$$

$$P_{Kspecies,error}(x) = \text{Poisson}(c_{error,out})$$

The $c_{error,out}$ and $c_{error,in}$ are related to the sequencing error rate, the k-mer size and the genome size， and they are quite difficult to be estimated in practice, so this sequencing error model was not used later.

To reduce the influence of erroneous k-mers, we suggest filtering k-mers by sequencing quality when counting the k-mer frequency. Here we calculate the probability that a k-mer is correct with the following formula:

$$P(base, error) = e_i = 10^{\frac{S}{-10}}$$

$$P(Kmer, correct) = \prod_{i=1}^{K} (1 - e_i)$$

Note: $S$ refer to the Phred scale score, which can be generated by most sequencing platforms. To filter most of the erroneous k-mers, we require all the k-mers being counted to have P(Kmer, correct) larger than a threshold, for example, 0.95.

## 7 Continuous compound model

Previous studies have shown that several factors may cause coverage bias, including uneven

GC content and specific sequence structure [7-9]. Then for real sequencing data, the coverage depth of sequenced k-mer is not only dependent on its genomic frequency, but also dependent on the sequencing bias characters, so we define the observed genomic frequency $\tilde{i}$ when sequencing coverage bias exists. The value of $\tilde{i}$ will be continuous. It is supposed that the sequencing on each group with same observed genomic frequency $\tilde{i}$ is independent, so for the k-mers from the same $\tilde{i}th$ group, the depth distribution is also following a Poisson distribution. Referring to the discrete compound model, we can get the continuous compound model to describe the total k-mer coverage depth distribution for real sequencing data:

$$P_{Kspecies}(x) = \int \tilde{a}_{\tilde{i}} \times P_{Kspecies,\tilde{i}}(x) d\tilde{i}$$

$$P_{Kindividuals}(x) = \int \tilde{b}_{\tilde{i}} \times P_{Kindividuals,\tilde{i}}(x) d\tilde{i}$$

Note: $\tilde{a}_{\tilde{i}}$ and $\tilde{b}_{\tilde{i}}$ are the observed value for each continuous $\tilde{i}$, which are different to the genomic $a_i$ and $b_i$. To make it easy, we used the following dense discrete model in practice:

$$P_{Kspecies}(x) = \sum_{k=1}^{m'} \tilde{a}_k \times P_{Kspecies,k}(x)$$

$$P_{Kindividuals}(x) = \sum_{k=1}^{m'} \tilde{b}_k \times P_{Kindividuals,k}(x)$$

Note: $m'$ is the total Poisson distribution number or peak number we considered. The symbol *k* here just stands for the peak order, and there is no relation between *k* and $\tilde{c}_k$. Then we estimate the $\tilde{c}_k$ and $\tilde{a}_k$ with the following iteration formulas:

$$P(j \mid k) = \frac{\tilde{c}_k{}^j e^{-\tilde{c}_k}}{j!}$$

$$\tilde{a}_{k,t+1} = \sum_{j=0}^{j=w} \frac{\tilde{a}_{k,t} P(j \mid k)}{\sum_k \tilde{a}_{k,t} P(j \mid k)} \times v_j$$

$$\tilde{c}_{k,t+1} = \frac{\sum_{j=0}^{w} j \times P(k \mid j) \times v_j}{\tilde{a}_{k,t+1}} = \frac{\sum_{j=0}^{w} j \times \frac{\tilde{a}_{k,t+1} P(j \mid k)}{\sum_k \tilde{a}_{k,t+1} P(j \mid k)} \times v_j}{\tilde{a}_{k,t+1}}$$

After we have estimated $\tilde{a}_k$ and $\tilde{c}_k$, we need to determine the coverage depth *c* and genomic $a_i$. Supposing that most of the genomic unique k-mers are not affected by coverage bias, so in practice, we can either use the $\tilde{c}_k$ with the highest $\tilde{a}_k$ in the unique peak region as the coverage depth *c*, or use the average $\tilde{c}_k$ around the highest $\tilde{a}_k$ to stand for the genomic coverage depth *c*. However, it is quite difficult to estimate the relationship between $\tilde{a}_k$ and $a_i$, because we can not model the sequencing coverage bias right now. Then in practice, we just calculate genomic $a_1$ by summarizing the $\tilde{a}_k$ values around the unique peak.

## 8 Dealing with sequencing error and coverage bias

To estimate the genome size by formula (2), we need to make clear the distribution of erroneous k-mers and estimate the parameters *n* and *c* for correct k-mers from the mixture k-mer coverage depth distribution. We have designed a united model to represent erroneous and correct k-mers together in this paper, however, the model is affected by many genomic and sequencing characters and thus quite difficult to resolve. Instead, we turned to use a simple experience based method to exclude the erroneous k-mers caused by sequencing errors. During each cycle of the Bayes iteration, the missing ratio values of k-mer species with depths lower than the threshold should also be re-calculated by the compound Poisson model with the input *c* and $a_i$ values like calculating $v_0$ in standard model.

Previous studies have shown that several factors may cause coverage bias, including uneven GC content and specific sequence structure [7-9]. In theory, although there may be some types of coverage bias that change the position of the major peak depth as well as the shape of the distribution curve, from our experience, the major peak depth from distribution curves with coverage bias still roughly reflects the expected coverage depth for the unique k-mer class (*c*) within most real sequencing data.

# Supplementary results

## 1 Estimating with various types of simulated data

We used ΔG/G to evaluate the accuracy of genome size estimation. Among the 96 analysis results, the largest ΔG/G value is 8.11%, which belongs to maize species with 1% heterozygous rate and 1% error rate and 25 k-mer size. We found that sequencing error is the most difficult factor to deal, because when the reads data has higher rate of sequencing error, it is more difficult to exclude the erroneous k-mers from the mixture distribution by the experience-based method. We also found the k-mer size influence the estimation accuracy. When there is sequencing error or heterozygosity, ΔG/G of 25-mer are relatively higher than that of 17-mer, which can be explained by that larger k-mer size will enhance the effect of heterozygosity and sequencing error, moreover, larger k-mer size also results in lower $c_{k\text{-}mer}$, which tends to decrease the estimation accuracy. Then at cases with no sequencing error or heterozygosity problems, larger k-mer size will generate more accurate estimation result by having more advantage to deal with repeat.

The evaluation of $a_i$ estimation and the analysis of influencing factors are much more complex than that of genome size. It is not only caused by the mixture effect of repeat and heterozygosity, but also related with the applied models (standard or heterozygous model). For simple, we used $\Delta a_1/a_1$ to evaluate the accuracy of $a_i$ estimation. Among the 96 analysis results, the largest $\Delta a_1/a_1$ value is 12.6%, which belongs to Maize species with 0.1% heterozygous rate and 1% error rate and 17 k-mer size.

## 2 Analyzing real sequencing data from *de novo* genome projects

By filtering the low-quality k-mers, there will be much less erroneous k-mers in the result, which not only makes it easier to estimate the total number of correct k-mer individuals but also significantly decrease the consumption of computer memory.

Taking the reported genome sizes in the published papers as reference, we calculated the ΔG/G values for the estimated genome sizes from various methods. About 22% and 80% of the ΔG/G values are smaller than 1% and 5% respectively, with one largest ΔG/G value of 10% for ant with all raw k-mers by rough estimation. Then we calculated the reference $a_1$ values from the assembled genome sequence and used it as a reference to calculate the $\Delta a_1/a_1$ values, among which 17% and 80% of the $\Delta a_1/a_1$ values are smaller than 1% and 5% respectively, with the largest $\Delta a_1/a_1$ value of 11% for ant low-quality filtered data with standard model.

There is observed difference between the estimated and reference $a_1$ ($b_1$) values. Besides the reasons described in the main text, the sequencing coverage bias and the uncertainty relationship between estimated $\tilde{a}_k$ values and genomic $a_i$ values should be the most influencing factor that caused this difference. Although the estimation accuracy is not as high as that of simulated data, the trends is still correct, and we can roughly infer that 15-mer is enough large to assemble Ecoli genome, whist larger than 25-mer is proper for assembling the human genome.

For real sequencing data with coverage bias, only when there is observable $c/2$ peak caused by heterozygosity, the heterozygous rate can be detected and roughly inferred. It is necessary to mention that the models in this paper are based on some special hypotheses, so for real sequencing data, when there is no clear peak which may be caused by low sequencing depth or high sequencing bias, the data is not qualified for genomic character estimation.

# References

1. Marcais G, Kingsford C: **A fast, lock-free approach for efficient parallel counting of occurrences of k-mers**. *Bioinformatics (Oxford, England)* 2011, **27**(6):764-770.
2. Melsted P, Pritchard JK: **Efficient counting of k-mers in DNA sequences using a bloom filter**. *BMC bioinformatics* 2011, **12**:333.
3. Wicker T, Sabot F, Hua-Van A, Bennetzen JL, Capy P, Chalhoub B, Flavell A, Leroy P, Morgante M, Panaud O *et al*: **A unified classification system for eukaryotic transposable elements**. *Nature reviews* 2007, **8**(12):973-982.
4. Benson G: **Tandem repeats finder: a program to analyze DNA sequences**. *Nucleic acids research* 1999, **27**(2):573-580.
5. She X, Cheng Z, Zollner S, Church DM, Eichler EE: **Mouse segmental duplication and copy number variation**. *Nature genetics* 2008, **40**(7):909-914.
6. Miller JR, Koren S, Sutton G: **Assembly algorithms for next-generation sequencing data**. *Genomics*, **95**(6):315-327.
7. Bentley DR, Balasubramanian S, Swerdlow HP, Smith GP, Milton J, Brown CG, Hall KP, Evers DJ, Barnes CL, Bignell HR *et al*: **Accurate whole human genome sequencing using reversible terminator chemistry**. *Nature* 2008, **456**(7218):53-59.
8. Nakamura K, Oshima T, Morimoto T, Ikeda S, Yoshikawa H, Shiwa Y, Ishikawa S, Linak MC, Hirai A, Takahashi H *et al*: **Sequence-specific error profile of Illumina sequencers**. *Nucleic acids research*.
9. Dohm JC, Lottaz C, Borodina T, Himmelbauer H: **Substantial biases in ultra-short read data sets from high-throughput DNA sequencing**. *Nucleic acids research* 2008, **36**(16):e105.